\documentclass{aa}
\usepackage{graphics,graphicx}
\usepackage[varg]{txfonts}
\usepackage{rotating}
\usepackage{lscape}
\usepackage{rotating}
\usepackage{natbib}
\usepackage{longtable}
\newcommand{\asec}{$^{\prime\prime}$}

\def\H{N$_{2}$H$^{+}$}

\def\CII{\mbox{C$^{18}$O}}

\def\kms{\mbox{km~s$^{-1}$}}
\def\cmc{cm$^{-3}$}
\def\cmq{cm$^{-2}$}

\def\solm{\mbox{$M_\odot$}}

\begin{document}
\title{The fragmentation properties of massive protocluster gas clumps: an ALMA study}
\author{F. Fontani\inst{1}
           \and
           B. Commer\c{c}on\inst{2} 
           \and
           A. Giannetti\inst{3}
           \and 
           M.T. Beltr\'an\inst{1}
           \and 
           \'A. S\'anchez-Monge\inst{4}
           \and 
           L. Testi\inst{1,5,6} 
           \and
           J. Brand\inst{7} 
           \and
           J.C. Tan\inst{6,8} 
           }
\offprints{F. Fontani, \email{fontani@arcetri.astro.it}}
\institute{INAF-Osservatorio Astrofisico di Arcetri, Largo E. Fermi 5, I-50125, Florence, Italy
          \and 
          Ecole Normale Sup\'erieure de Lyon, CRAL, UMR CNRS 5574, Universit\'e Lyon I, 46 All\'ee d'Italie, 69364, Lyon Cedex 07, France
          \and
          Max-Planck-Institut f\"{u}r Radioastronomie, auf dem H\"{u}gel 69, 53121 Bonn, Germany
          \and
          I. Physikalisches Institut, Universit\"{a}t zu K\"{o}ln, Z\"{u}lpicher Str. 77, 50937 K\"{o}ln, Germany
          \and
          European Southern Observatory, Karl-Schwarzschild-Str 2, D-85748, Garching bei M\"{u}nchen, Germany
          \and 
          Gothenburg Center for Advance Studies in Science and Technology, Chalmers University of Technology and University of Gothenburg, SE-412 96 Gothenburg, Sweden
          \and
          INAF-Istituto di Radioastronomia and Italian ALMA Regional Centre, via P. Gobetti 101, I-40129, Bologna, Italy
          \and
          Department of Astronomy \& Physics, University of Florida, Gainesville, FL 32611, USA
          }
\date{Received date; accepted date}

\titlerunning{Magnetically regulated fragmentation}
\authorrunning{Fontani et al.}


\abstract{Fragmentation of massive dense molecular clouds is the starting point
in the formation of rich clusters and massive stars. Theory and numerical simulations 
indicate that the population of the fragments (number, mass, diameter, separation) resulting
from the gravitational collapse of such clumps is probably regulated by the 
balance between the magnetic field and the other competitors of self-gravity,
in particular turbulence and protostellar feedback.
We have observed 11 massive, dense and young star-forming clumps with the Atacama
Large Millimeter Array (ALMA) in the thermal dust continuum emission at $\sim 1$~mm 
with an angular resolution of $0\farcs25$ with the aim of determining their 
population of fragments. The targets have been selected from a sample of 
massive molecular clumps, with limited or absent star formation activity, and hence
limited feedback. We find fragments on sub-arcsecond 
scales in 8 out of the 11 sources. The ALMA images indicate two different fragmentation modes: 
a dominant fragment surrounded by companions with much smaller mass 
and size, and many ($\geq 8$) fragments with a gradual change in masses and sizes.
The morphologies are very different, with three sources that show filamentary-like
distributions of the fragments, while the others have irregular geometry.  
On average, the largest number of fragments is found towards the warmer and more 
massive clumps. Also, the warmer clumps tend to form 
fragments with larger mass and size. To understand the role of the different physical
parameters to regulate the final population of the fragments, we have simulated the 
collapse of a massive clump of $100$ and $300$ \solm\ having different magnetic
support. The 300 \solm\ case has been run also for different initial temperatures 
and Mach numbers $\mathcal{M}$ to evaluate the separate role of each of these
parameters. The simulations indicate that: (1) fragmentation is inhibited when the
initial turbulence is low ($\mathcal{M}\sim 3$), independent of the other physical parameters. 
This would indicate that the number of fragments in our clumps can be explained 
assuming a high ($\mathcal{M}\sim 6$) initial turbulence, although an initial density
profile different to that assumed can play a relevant role; 
(2) a filamentary distribution of the fragments is favoured in a highly magnetised clump. 
We conclude that the clumps that
show many fragments distributed in a filamentary-like structure are likely characterised
by a strong magnetic field, while the other morphologies are possible also in a weaker 
magnetic field.}


\maketitle

%

\section{Introduction}
\label{intro}

Massive and dense molecular clumps (compact structures with $M\geq 100$ \solm , and 
$n$(H$_2$)$\geq 10^4$\cmc ) in infrared-dark clouds are believed to be the birthplaces of 
rich clusters and high-mass O-B stars (e.g.~Ragan et al.~\citeyear{ragan2011}; Peretto 
et al.~\citeyear{peretto}; Tan et al.~\citeyear{tan2013}; Rathborne et al.~\citeyear{rathborne}).
The formation of these systems starts with the fragmentation of the parent clump occuring
during its gravitational collapse, which is thus a crucial process in determining the final stellar 
population. In particular, the process has important implications in the theoretical debate of 
massive star formation ($M_{*}\geq 8$ \solm ), because the two main competing theories
assume a totally different degree of initial fragmentation: in the core-accretion models 
(e.g. McKee \& Tan~\citeyear{met}), massive stars
are born from the direct collapse of a near-equilibrium clump in which only one (or very few)
fragments form; in the competitive accretion models (e.g.~Bonnell et al.~\citeyear{bonnell2004}),
the parent clump fragments into many low-mass seeds of the order of the thermal Jeans mass
which competitively accrete from the common unbound gaseous envelope.

Theoretical models and simulations predict that the number, the size, the mass, and the spatial 
distribution of the fragments depend strongly on which of the main competitors of gravity is dominant. 
The main physical mechanisms that oppose gravity during collapse are: thermal pressure, 
intrinsic turbulence, protostellar feedback (such as outflows or expanding H$_{\rm II}$ regions), 
and magnetic pressure (e.g.~Krumholz~\citeyear{krumholz}, 
Hennebelle et al.~\citeyear{hennebelle}, Federrath et al.~\citeyear{federrath2015}). 
However, at the beginning of the gravitational collapse,
the thermal support is expected to be negligible. Mechanical feedback from nascent 
protostellar objects through outflows and jets, expected to be launched early in the evolution 
of protostars (Krumholz et al.~\citeyear{krumholz2014}), can affect the earliest phases of
the fragmentation process (Federrath et al.~\citeyear{federrath2014}), especially from newly 
born massive objects. Other feedback such as powerful stellar winds or expanding HII regions
are expected to appear only in evolved stages and should not influence early fragmentation
(Bate~\citeyear{bate}).
Therefore, the fragmentation at the earliest stages is influenced mainly by magnetic support, 
intrinsic turbulence, and protostellar feedback. However, in objects with no observational
evidence of protostellar outflows, the contribution of protostar feedback to fragmentation 
should not dominate, and the fragment population should be 
mostly due to the competition between magnetic field and intrinsic turbulence.
In this respect, Commer\c{c}on et al.~(\citeyear{commercon2011})
have shown that if the magnetic support dominates the dynamical evolution, only one
(or few) fragments surrounded by a non-fragmenting envelope are expected, while many
small fragments with mass of the order of $\sim 0.1 - 1$ \solm\ separated by projected 
distances of $\sim 100-1000$ au are foreseen if the magnetic support is weak.

An understanding of the formation of massive stars and rich clusters thus requires observational
studies of massive dense cores in a very early stage of evolution, with both sensitivity and
angular resolution appropriate to detect and resolve the smallest fragments predicted by the
simulations. Surveys of massive dense clumps with adequate resolution (of the order of 
$\simeq 0.1-1$ \asec, corresponding to $\sim 100 - 1000$ au at 1 kpc) and sensitivity 
(of the order of $\simeq 0.1$ \solm ) reveal either a few fragments (e.g.~Bontemps et 
al.~\citeyear{bontemps}, Longmore et al.~\citeyear{longmore}, Palau et al.~\citeyear{palau}, 
Csengeri et al.~\citeyear{csengeri}), or structures with large (ten or more) number of 
fragments (e.g.~Zhang et al.~\citeyear{zhang}, Rathborne et al.~\citeyear{rathborne}, 
Palau et al.~\citeyear{palau2017}, 
Henshaw et al.~\citeyear{henshaw2017}, Cyganowski et al.~\citeyear{cyganowski}). 
In regions with many fragments, the interpretation of existing studies is very complex: 
in some cases, the properties of the fragments do not seem consistent with 
a pure gravo-turbulent scenario (e.g.~Zhang et al.~\citeyear{zhang}), but
in others they can be explained with a pure thermal Jeans fragmentation (Palau et al.~\citeyear{palau2015},
Palau et al.~\citeyear{palau2017}), or they seem to belong to complex sub-structures
difficult to explain with simple theoretical models (e.g.~Henshaw et al.~\citeyear{henshaw2017},
Cyganowski et al.~\citeyear{cyganowski}). 
These results indicate that non-thermal forms of energy could play a relevant role in regulating 
the fragmentation at these small scales, but, overall, to date no firm conclusions can be
derived. 

In this work, we present an ALMA survey of 11 massive 
dense clumps in the thermal dust continuum emission at $\sim 278$~GHz with angular resolution 
$0\farcs 25$, and mass sensitivity of the order of $\sim 0.1$~\solm , or better.
In the first source belonging to this survey studied in detail, 
16061--5048c1 (Fontani et al.~\citeyear{fontani2016}),
we have detected 12 fragments, most of them located in a filament-like structure
coincident with the location of an embedded 24~$\mu$m source.
Although at first glance the large number of fragments could indicate a fragmentation
process induced by a faint magnetic support, simulations run specifically for this object, i.e.
assuming as initial conditions (temperature, mass, and Mach number) those of this source
obtained from previous observations, suggest that instead its fragment population 
can be explained better with a strong magnetic support, especially because the
filament-like morphology detected cannot be obtained with a faint magnetic
support.
The goal of the present work is to expand the study of 16061--5048c1 to a larger 
sample of objects selected similarly, in order to better understand the dominant 
ingredient regulating the fragmentation process in collapsing massive dense clumps
in very early stages of evolution. 
In Sect.~\ref{sample} we present the source sample and the criteria
used to select it; Sect.~\ref{obs} describes the observations, and Sect.~\ref{res}
the observational results; in Sect.~\ref{discu}, we discuss our findings 
based on the help of numerical simulations. Finally, in Sect.~\ref{conc} 
we give a brief summary of our work, and draw the most relevant conclusions.

\section{Source sample}
\label{sample}

The targets have been selected from an initial sample of MSX-dark clumps 
(Beltr\'an et al.~\citeyear{beltran2006}) detected at 1.2~mm with the SIMBA
bolometer at the SEST. The selection criteria applied make us confident
that all objects are: (1) potential sites of massive star formation, (2) dense, 
(3) quiescent, (4) cold and chemically 
young. To satisfy these criteria, we selected clumps having the following observational 
properties: (1) gas mass and gas surface (and column) density consistent 
with being potential sites of massive star formation according to observational 
findings (Kauffmann \& Pillai~\citeyear{kep}); (2) detection in the high-density gas 
tracer \H\ (3--2) with APEX (Fontani et al.~\citeyear{fontani2012}), which is also
the most reliable tracer of dense molecular gas (Kauffmann et al.~\citeyear{kauffmann2017}); 
(3) clumps isolated or having the 1.2~mm emission peak well separated ($\geq 24$\asec) from that 
of other clumps, and without evidence of star formation activity (Beltr\'an et al.~\citeyear{beltran2006}, 
S\'anchez-Monge et al.~\citeyear{sanchez}); (4) average CO depletion factor (ratio between 
expected and observed CO abundance) derived from APEX observations of \CII\ (3--2), 
$f_{\rm CO} \geq 7$ (Fontani et al.~\citeyear{fontani2012}), which provides evidence of 
the chemical youth of the clumps. Clump coordinates, distances, and main physical 
properties of the 11 selected clumps are summarised in Table~\ref{tab_simba}. 

\begin{figure*}[!]
\centerline{\includegraphics[width=16cm,angle=0]{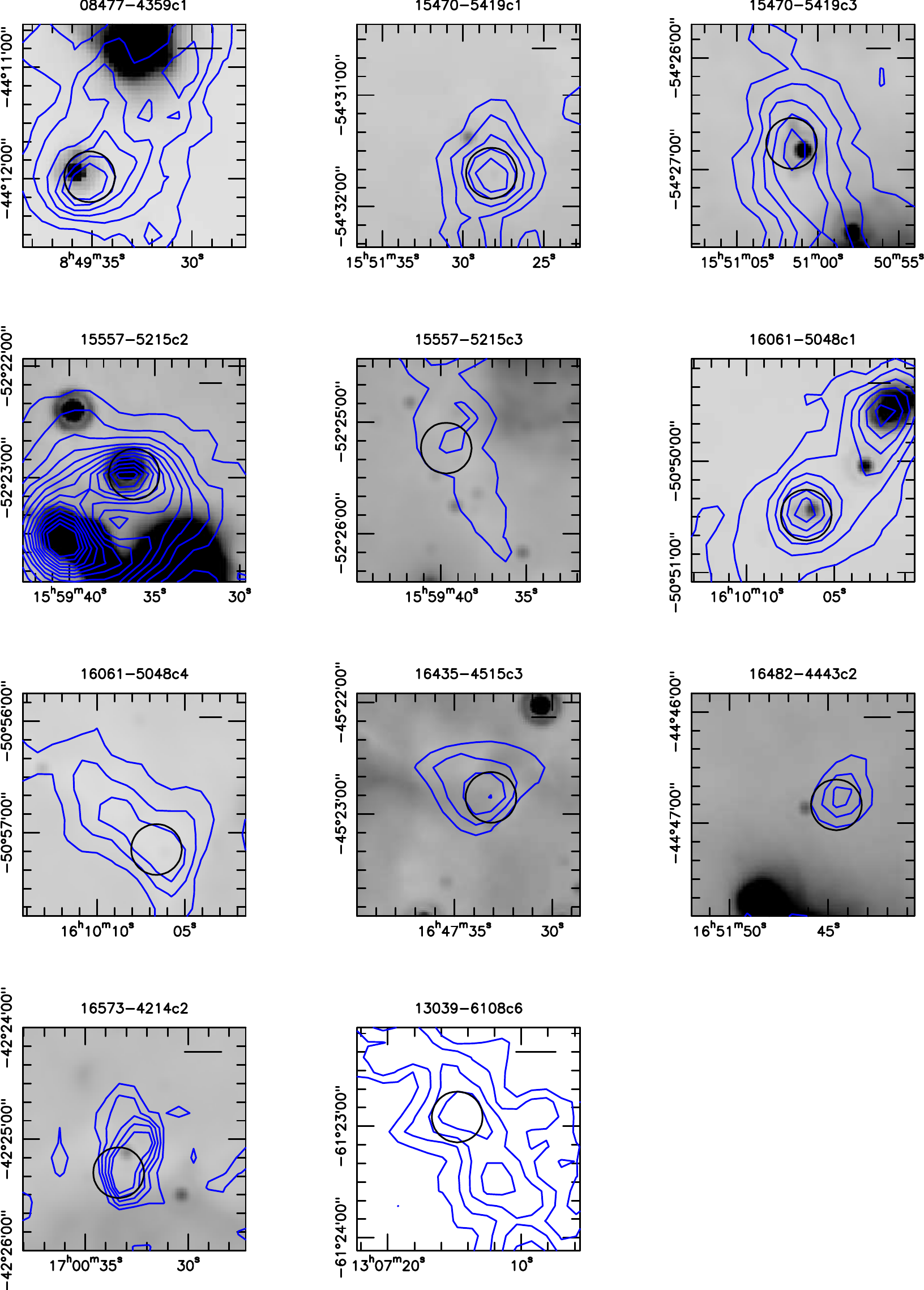}}
\caption{1.2~mm continuum maps (contours) obtained with SIMBA at the SEST towards the 11 sources 
in Table~1, having an angular resolution of $\sim$24\asec . In each panel, the image on the background is the 
Spitzer-MIPS 24~$\mu$m map, available for all clumps but 13039--6108c6, and the circle indicates the ALMA field 
of view at the frequency of the \H\ (3--2) line ($\sim$22\asec ) centred at the coordinates given in 
Table~\ref{tab_simba}. The first contour and step correspond to the $3\sigma$ rms level in the map,
with the exception of 16061--5048c1, in which the step is of $6\sigma$ rms (see Beltr\'an et al.~\citeyear{beltran2006}
for details). In each panel, the horizontal black bar in the top right corner shows a linear scale of 0.25~pc.
}
\label{fig_simba}
\end{figure*}

The 1.2~mm continuum maps of all clumps are shown in Fig.~\ref{fig_simba}, 
superimposed on the Spitzer 24 $\mu$m images.
Some of the clumps are detected at 24 $\mu$m, which indicates a potential on-going
star formation activity. However, the observational selection criteria (3) and (4)
make us confident that possible embedded protostellar activity has not affected 
significantly the environment yet. Therefore, outflows, jets or other
forms of mechanical protostellar feedback should not dominate in determining the
fragment population.

The young evolutionary stage of the sources is also strongly supported by their low
Star Formation Efficiency (SFE), given in the last Column of Table~\ref{tab_simba}. 
The SFE has been calculated according to:
\begin{equation}
{\rm SFE} = M_{\rm stars}/(M_{\rm gas} + M_{\rm stars})\;,
\end{equation}
where $M_{\rm stars}$ is the mass already in the form of (proto-)stars calculated from
the source bolometric luminosity (Giannetti et al.~\citeyear{giannetti}) following the
approach in Beltr\'an et al.~(\citeyear{beltran2013}), and $M_{\rm gas}$ is the gas mass listed
in Table~\ref{tab_simba} derived by Giannetti et al.~(\citeyear{giannetti}) using the 
dust thermal continuum emission in Beltr\'an et 
al.~(\citeyear{beltran2006}). $M_{\rm stars}$ is computed from the bolometric luminosity
assuming that the infrared emission is consistent with that of an embedded stellar cluster, 
although care needs to be taken due to the contribution from accretion luminosity. 
This caveat is especially relevant taking into account the fact that most stars should 
be of low mass, for which the accretion luminosity is expected to dominate.
SFE is below 20$\%$ in all targets but 16061--5048c1, for which SFE is $\sim 31\%$.

\begin{table*}[tbh]
\begin{center}
\caption[]{Sample of massive dense clumps and general properties: coordinates, distance, 
deconvolved angular diameter, gas mass, gas temperature, H$_2$ column density, mass surface 
density, CO depletion factor, and non-thermal velocity dispersion (parameters taken or derived
from Beltr\'an et al.~\citeyear{beltran2006}, Giannetti et al.~\citeyear{giannetti}, and Fontani
et al.~\citeyear{fontani2012}). In the last two columns, we give the recovered flux in the 
ALMA images, and the Star Formation Efficiency (SFE), computed as explained in Sect.~\ref{sample}.}
\label{tab_simba}
\tiny
\begin{tabular}{lcccccccccccc}
\hline \noalign {\smallskip}
Source  & R.A.;Dec.(J2000) & $l$;$b$ & $d$$^{(a)}$ & $\theta_{\rm s}$$^{(a)}$ &  $M_{\rm gas}$$^{(b)}$ & $T_{\rm k}$$^{(b)}$ & $N({\rm H_2})$$^{(b)}$ & $\Sigma({\rm H_2})$$^{(b)}$ & $f_{\rm CO}$$^{(c)}$ & $\sigma_{\rm nth}$$^{(d)}$ & Rec. flux$^{(e)}$ & SFE \\
     & ${\rm h\,m\,s}$;$\circ\,\prime\,\prime\prime$ & $\circ$;$\circ$ &  kpc & \asec\ & \solm\ & K & $10^{23}$ \cmq\ & g \cmq\ & & \kms\ & Jy/Jy & $\%$ \\
\hline \noalign {\smallskip}
08477--4359c1$^{(f)}$ &  08:49:35.13;--44:11:59 & 264.69;--0.07  & 1.8 & 	35.6   & 86.73 &  19   &   1.42   & 0.24  &    7 & 1.03 & 0.12/0.62 & --$^{(g)}$ \\ 
13039--6108c6  & 13:07:14.80;--61:22:55  & 305.18;1.14  & 2.4 & 	40.3 &   101.5 &  17   &   0.68  & 0.12   &    22 & -- & 0 & --$^{(g)}$ \\
15470--5419c1   & 15:51:28.24;--54:31:42 & 327.51;--0.83 &   4.1 &   24.2     &  310.2 & 18  &   1.37    & 0.36 &    35 & 1.02 & 0.01/0.56 & $6\%$ \\
15470--5419c3$^{(f)}$   & 15:51:01.62;--54:26:46 & 327.51;--0.72  &  4.1 &   54.1      &  743.4 & 19 &   1.11    & 0.17 &  36 & 1.13 & 0.09/0.50 & $3\%$ \\  
15557--5215c2$^{(f)}$   & 15:59:36.20;--52:22:58 & 329.81;0.03 &  4.4 &   41.3     &  633.4 &  23 &   1.55    & 0.22 &   32 & 0.96 & 0.12/0.90 & $4\%$ \\
15557--5215c3   & 15:59:39.70;--52:25:14  & 329.80;0.00 &  4.4 &   35.8      &  194.3 &  15 &   0.49    & 0.09 &   24 & -- &  0 & $8\%$ \\
16061--5048c1$^{(f)}$   & 16:10:06.61;--50:50:29  & 332.06;0.08 &  3.6 &   28.1      & 284.3 &  25 &   1.66    & 0.31 &   12 & 1.52 & 0.63/1.02 & $31\%$ \\
16061--5048c4   & 16:10:06.61;--50:57:09  & 331.98;0.00 &   3.6 &   62.8     &   504.2 & 13 &   1.22    & 0.11 &   34 & 0.82 & 0.03/0.32 & $3\%$ \\ 
16435--4515c3   & 16:47:33.13;--45:22:51  & 340.31;--0.71 &  3.1 &  17.7    &	147    &   12  &   1.20   & 0.55  &  73 & -- & 0 & $11\%$  \\
16482--4443c2   & 16:51:44.59;--44:46:50  & 341.24;--0.90 & 3.7 &  $\ll 24$$^{h}$ & 59.08 & 16 & $\gg 4.63^{h}$ & 0.66 &   9 & 1.40 & 0.07/0.23 & $17\%$ \\
16573--4214c2   & 17:00:33.38;--42:25:18  & 344.08;--0.67 &  2.6 &  7.29  &	108.3  &  17 &	   1.89   & 3.4   &  25  & 1.17 & 0.07/0.71 & $14\%$ \\
\hline \noalign {\smallskip}
\end{tabular}
\end{center}
$^{(a)}$ from Beltr\'an et al.~(\citeyear{beltran2006}); \\
$^{(b)}$ from Giannetti et al.~(\citeyear{giannetti}); \\
$^{(c)}$ from Fontani et al.~(\citeyear{fontani2012}); \\
$^{(d)}$ derived from the \CII\ (3--2) line width at half maximum (Fontani et 
al.~\citeyear{fontani2012}) by subtracting the thermal contribution calculated according 
to the gas temperature in Col.~7; \\
$^{(e)}$ ratio between the total flux integrated inside the ALMA primary beam, and the
peak flux of the SIMBA map towards the phase centre. Please note that the SIMBA
main beam and ALMA primary beam are the same ($\sim 24$\asec ), and that the
flux ratios have been compared by correcting the SIMBA flux at 250~GHz assuming
a spectral index $\beta=2$; \\
$^{(f)}$ detected in the Spitzer 24~$\mu$m image (Fig.~\ref{fig_simba}); \\
$^{(g)}$ not possible to derive SFE because the bolometric luminosity is not available (Giannetti et al.~\citeyear{giannetti}); \\
$^{(h)}$ point-like source in the SIMBA 1.2~mm map (Beltr\'an et al.~\citeyear{beltran2006}). \\
\end{table*}
\normalsize

\section{Observations and data reduction}
\label{obs}

Observations with the ALMA array at a frequency of $\sim 278$~GHz were performed during cycle-2 
and 3 in configuration C36-6 with baselines up to 1091 m, providing an angular resolution of 
$0\farcs 25$, and a maximum recoverable scale of $3\farcs 5$. For each clump, 
the phase centre was set to the coordinates given in Table~\ref{tab_simba}. The total integration 
time on each source was $\sim 18 - 20$ minutes. The amount of precipitable water vapour 
during observations was generally around $\sim 1.5 - 2$ mm. Bandpass and phases were 
calibrated by observing J1427--4206 and J1617--5848, respectively. 
The absolute flux scale was set through observations of Titan and Ceres.
Continuum was extracted by averaging in frequency the line-free channels. 
The total bandwidth used is $\sim 1.7$~GHz.
Calibration and imaging were performed with the CASA\footnote{The Common Astronomy 
Software Applications (CASA) software can be downloaded at http://casa.nrao.edu} 
software (McMullin et al. 2007). Primary beam correction was always applied, 
and the final images were analysed following standard procedures with the 
software MAPPING of the GILDAS\footnote{http://www.iram.fr/IRAMFR/GILDAS} 
package. The angular resolution of the final images is $\sim 0\farcs 25$ . 
We are sensitive to unresolved fragments of $\gtrsim 0.05 - 0.1$ \solm .
We estimated the missing flux by comparing the total integrated
flux in the primary beam of the ALMA images with the single-dish
continuum measured by Beltr\'an et al.~(\citeyear{beltran2006}). 
The ratios are given in Table~\ref{tab_simba}.

\section{Results}
\label{res}

The ALMA maps of the dust thermal continuum emission, corrected for the primary
beam, are shown in Fig.~\ref{fig_maps_tot}. The plot aims to compare the morphology 
of the fragment population in the targets to understand possible global similarities 
and differences. The same images, with the fragment identification and
a better presentation of the emission morphology in each source, is given in 
Appendix~A. 

The dust thermal continuum emission has been decomposed into fragments according
to the following criteria: (1) peak intensity greater than 5 times the noise level; 
(2) two partially overlapping fragments are considered as resolved if they 
are separate at their half peak intensity level. 
The minimum threshold of 5 times the noise was adopted according to the fact 
that some peaks at the edge of the primary beam are comparable to about 4-5 
times the noise level. We decided to use these criteria and decompose the map 
into cores by eye instead of using decomposition algorithms (such as Clumpfind) 
because small changes in their input parameters could lead to big changes in 
the number of identified clumps (Pineda et al.~\citeyear{pineda}).
In Appendix-A, from Fig.~\ref{fig_map_08477} to Fig.~\ref{fig_map_16573}, we show the 
fragments identified in each source superimposed on the corresponding ALMA
continuum image. We refer to Fig.~1 of Fontani et al.~(\citeyear{fontani2016}) for 
the map of 16061--5048c1, with the identified fragments.

We have detected fragments in eight out of the 11 targets observed, 
and found at least four significant fragments in all of the eight clumps (see 
Figs.~\ref{fig_map_08477} -- \ref{fig_map_16573} in Appendix~A for a detailed 
description). Towards 13039--6108c6, 15557--5215c3, and 16435--4515c3 we do 
not detect any significant (peak flux $\geq 5\sigma$ rms) fragment. The maps of these
sources are shown in Fig.~\ref{fig_maps_nofragments}. This indicates 
that the emission is either more extended than the maximum recoverable 
angular scale ($\sim 3.5$\asec), so that we totally resolve it out, or that
the phase centre is not located at the actual centre of the fragmenting region. 
An uncertainty in the position of the phase centre can influence both the detection
and the number of fragments and the amount of missing flux. 
We will discuss this point further in Sect.~\ref{res_morf}. 

\subsection{Morphology of the continuum emission}
\label{res_morf}

\begin{figure*}[!]
\centerline{\includegraphics[width=18cm,angle=0]{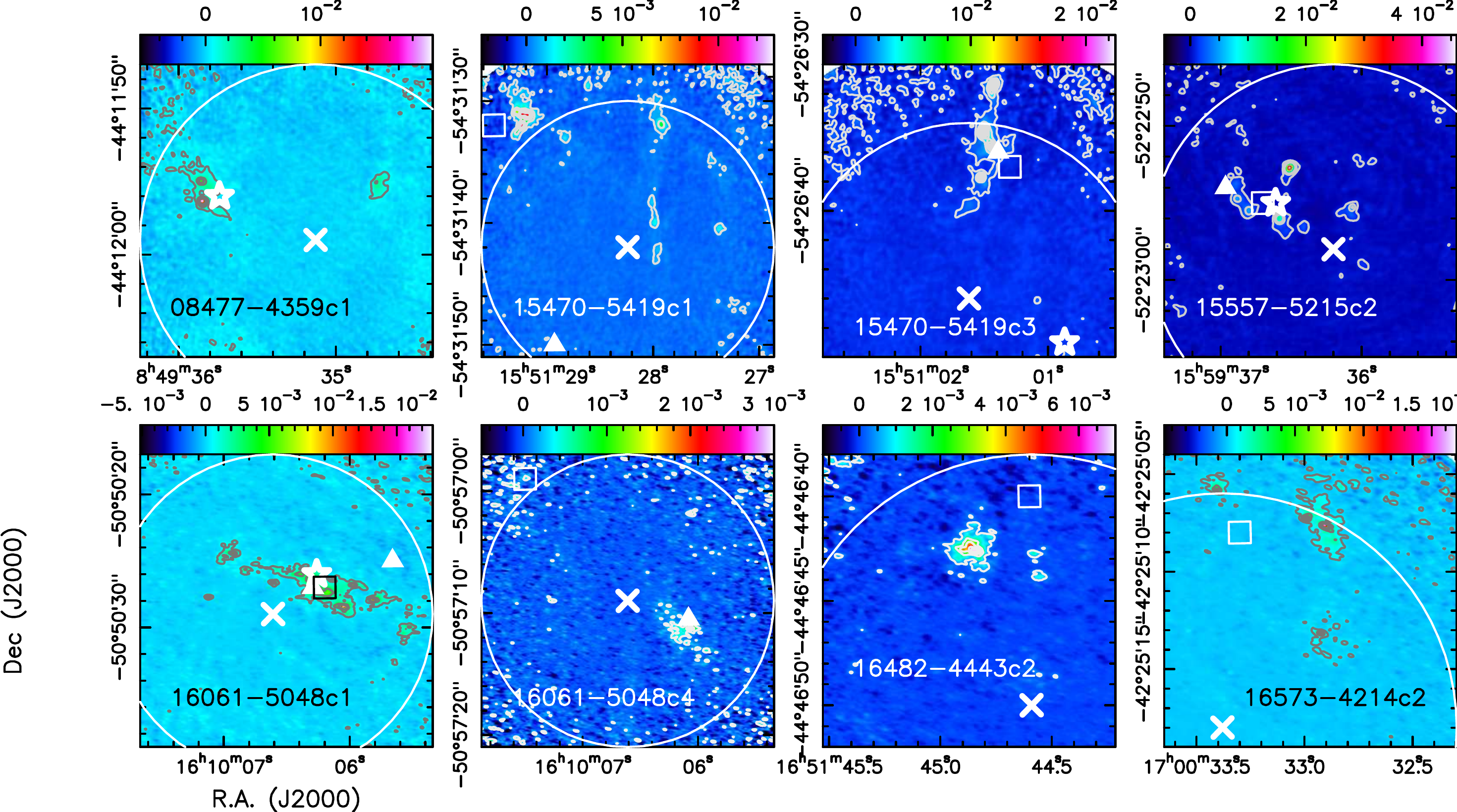}}
\caption{Dust thermal continuum emission maps (contours) at 278~GHz obtained with 
ALMA with angular resolution of $\sim 0\farcs 25$ towards the eight targets detected.
All images are primary beam corrected.
The wedge on top of each panel indicates the flux density scale (in Jy beam$^{-1}$). 
The target names are reported at the bottom of each frame. 
Three targets were observed but undetected: 13039--6108c6, 15557--5215c3, 
and 16435--4515c3. Their maps are shown in Fig.~\ref{fig_maps_nofragments}.
Contours start from the 3$\sigma$ rms level, and are in steps of 
10--20$\sigma$ rms, depending on the source. In each panel, the white circle indicates the ALMA field 
of view at 278~GHz ($\sim$24\asec ) centred on the single-dish mm continuum peak marked
by the cross (Beltr\'an et al.~\citeyear{beltran2006}). The white stars show the eventual
Spitzer 24~$\mu$m continuum peak detected in the ALMA field of view (see Fig.~\ref{fig_simba}),
and the filled triangles pinpoint the position of the H$_2$O maser spots detected towards
some clumps (Giannetti et al.~\citeyear{giannetti}). The square shows the emission peak
detected in ATLASGAL, at $\sim 870$ $\mu$m (Schuller et al.~\citeyear{schuller2009};
source 08477--4359c1 is not present in the ATLASGAL catalogue).
}
\label{fig_maps_tot}
\end{figure*}

The number of fragments detected ranges from a minimum of four to a maximum
of 14 fragments. 
Fig.~\ref{fig_maps_tot} presents all the detected sources. The morphologies are very 
different, with three sources in which the fragments are located along a 
filament-like structure, 15470--5419c3, 15557--5215c2, and 16061--5048c1, 
while the others have irregular geometry. About the relative intensity of the 
fragments within each clump, one can roughly distinguish between sources with a 
dominant fragment, like 15470--5419c1, 15470--5419c3, 15557--5215c2, and 
16482--4443c2, and objects with a smoother distribution in intensity of the fragments. 
The presence of a dominant fragment can be understood from the average mass 
ratio between the more massive fragment and the others (Col.~6 in 
Table~\ref{tab_averages}): for the four clumps mentioned above, this ratio is
larger ($\geq 18$) than for the others ($\leq 10$). The fragment mass, 
$m$, has been calculated following Eq.~(A1) in Fontani et al.~(\citeyear{fontani2016}),
taking the clump distances in Table~\ref{tab_simba}, and assuming as dust temperature 
the average clump gas temperature listed in Table~\ref{tab_simba}. 
This latter assumption is critical, because some fragments probably have higher 
temperatures, especially those associated with the 24~$\mu$m emission, in which 
the star formation activity is expected to be higher (i.e.~08477--4359c1, 15470--5419c3, 
15557--5215c2, and 16061--5048c1).
In these cases, our mass estimates are likely upper limits. This issue can be
solved only with a high angular resolution map of the dust temperature, unavailable
to date. Finally, we have assumed the same gas-to-dust ratio (100) and the same
expression for the dust mass opacity index as in Fontani et al.~(\citeyear{fontani2016}). 
The errors on the gas masses calculated in this way are difficult
to quantify, mostly because of the large uncertainty in the mass 
opacity coefficient, which can be up to a factor 2-3 (e.g.~Ossenkopf \& Henning~\citeyear{oeh}).

Four objects have Spitzer 24 $\mu$m emission
(indicated by the star in Fig.~\ref{fig_maps_tot}) within the primary beam, 
and in three of them, 08477--4359c1, 15557--5215c2, and 16061--5048c1, 
the fragments are clearly associated with the infrared source. 
The only exception is 15470--5419c3, for which the fragments
appear totally offset from both the Spitzer source and the phase centre, at the
border of the ALMA primary beam. This morphology, however, is in rough
agreement with the elongated structure seen in the SIMBA map.

\begin{figure*}[!]
\centerline{\includegraphics[width=18cm,angle=0]{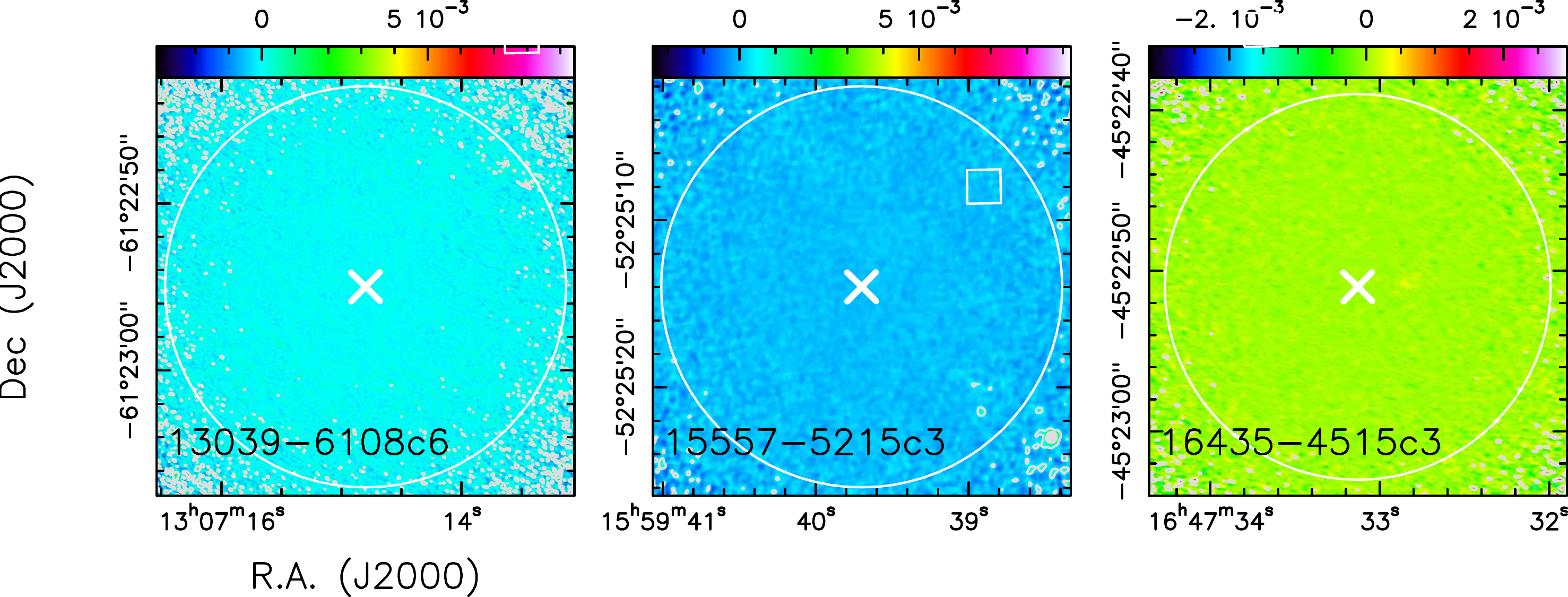}}
\caption{Same as Fig.~\ref{fig_maps_tot} for the three objects undetected with
ALMA: 13039--6108c6, 15557--5215c3, and 16435--4515c3. The location of
the ATLASGAL peak is just outside the field shown in 13039--6108c6 and 16435--4515c3.}
\label{fig_maps_nofragments}
\end{figure*}

None of the fragments coincides with the peak of the emission as
mapped by SIMBA (indicated by the crosses in Fig.~\ref{fig_maps_tot}). 
In general, the
asymmetric location of the fragments with respect to the phase centre is in
rough agreement with the asymmetric emission seen with the single-dish 
(see Fig.~\ref{fig_simba}), but larger than the nominal pointing error 
(estimated to be of a few arcseconds, Beltr\'an et al.~\citeyear{beltran2006}).
The exception is 16482--4443c2, in which the fragments are located to the
North-East, while the SIMBA map seems rather to be slightly elongated to the
West (although the SIMBA source is considered as unresolved
by Beltr\'an et al.~\citeyear{beltran2006}). We have checked if this can be 
due to a larger SIMBA pointing uncertainty by comparing the ALMA maps with the ATLASGAL
images (Schuller et al.~\citeyear{schuller2009}) at $\sim 870$ $\mu$m. In fact, 
because both the observing frequency and the angular resolution of ATLASGAL
are similar to those of our SIMBA data, but have lower noise level, the
ATLASGAL maps can help us to pinpoint the single-dish emission peak
with better signal-to-noise ratio. All our clumps but 08477--4359c1 are present in 
the ATLASGAL catalogue. The emission peak of the $\sim 870$ $\mu$m images is 
superimposed on the ALMA images in Figs.~\ref{fig_maps_tot} and~\ref{fig_maps_nofragments}: 
indeed, in most of the detected sources the ATLASGAL emission peak is more 
consistent with the location of the ALMA fragments, and offset from the SIMBA 
peak by a comparable angular displacement. In particular, Fig.~\ref{fig_maps_tot} 
shows that the angular separation between the SIMBA and ATLASGAL 
peaks is in between 3\asec\ for 16061--5048c1 and 13\asec\ for 15470--5419c1. The clumps in 
which the separation is the largest are 15470--5419c1, 16573--4214c2, and 16061--5048c4,
but those in which the effect is most important are 15470--5419c1, 15470--5419c3, and
16573--4214c2, because several intense fragments appear to be located at the border
of (or even outside) the primary beam. Hence, in these sources the number of the 
fragments and the recovered flux have to be considered as lower limits. 
The fragments identified outside the primary beam have been considered 
significant and included in the analysis only if their intensity peak is $\geq 10\sigma$ rms,
to avoid fake detections due to the worse signal-to-noise ratio at the edge of
the maps.

Among the undetected sources, the ATLASGAL emission peak is outside the primary beam
in 13039--6108c6 and 16435--4515c3. Therefore, probably the non-detection of 
fragments towards these two objects is due to the SIMBA pointing error.
On the other hand, in 15557--5215c3 the ATLASGAL peak is offset with respect to the SIMBA peak
only by 7\asec, so well inside the ALMA primary beam. We propose that the lack of fragments 
in this source could be due to the fact that the emission is extended and not (yet) distributed
into dense and compact condensations. 
The absence of embedded infrared sources and the relatively low ($15$~K) gas 
temperature are consistent with the very early evolutionary stage of this source. 

The maps of 15470--5419c1 and 16061--5048c4 require an additional comment: ALMA reveals 
several fragments in both regions, but the missing flux is huge ($98\%$
in 15470--5419c1, and $90\%$ in 16061--5048c4, see Table~\ref{tab_simba}). 
This latter has been obtained from the ALMA images by comparing the flux density 
integrated in the primary beam ($\sim 24$\asec ) to the peak flux of the SIMBA map 
(given that the SIMBA beam is also $\sim 24$\asec ).
However, in both sources the ATLASGAL emission peak is just outside the primary beam.
In particular, in 16061--5048c4 the morphology of the detected feature resembles that of 
an extended object elongated in direction NE-SW, in which the fainter fragments around 
the main one could be residuals of the envelope partially resolved out, and not real dust 
condensations (see also Fig.~\ref{fig_map_16061}).
All this makes any interpretation of the fragment population in 
16061--5048c4 very uncertain. The same comment applies to 15470--5419c1,
in which the interpretation of the fragment population must be taken with
caution because the location of the most massive fragments are at the border of the 
ALMA primary beam.


\subsection{Physical properties of the fragments}
\label{res_prop}

In Appendix-A, from Table~\ref{tab_08477} to Table~\ref{tab_16573}, we list the 
main properties of the fragments: peak position, integrated flux 
density ($F_{\nu}$), peak flux density ($F_{\nu}^{\rm peak}$), diameter ($D$), 
and mass ($m$). To derive these parameters, we adopt the same approach as in 
Fontani et al.~(\citeyear{fontani2016}), hence we give in the following a brief description 
of the methods adopted to compute them, and we refer to Sect.~A.1 of that paper for any 
other detail. We also refer to the same paper for the properties of the fragments
identified in 16061--5048c1 (Fig.~1 and Appendix-A of Fontani et al.~\citeyear{fontani2016}).

For each fragment, $F_{\nu}$ has been computed by integrating the flux density inside the white
polygon depicted in Figs.~\ref{fig_map_08477}$-$\ref{fig_map_16573}, which corresponds 
to the 3$\sigma$ rms level of the map. In the cases in which the 3$\sigma$ level
of two adjacent fragments were not separate, the edges between the two have 
been defined by eye at approximately half of the separation between the peaks.
The diameter, $D$, of each fragment has been computed as the diameter of the circle
having the same surface of the fragment. Finally, the fragment mass, $m$, has been 
calculated as explained in Sect.~\ref{res_morf}.

The physical properties of the fragments found in each source, calculated following 
the aforementioned methods, are shown in Tables~\ref{tab_08477} -- \ref{tab_16573}.
In Table~\ref{tab_averages} we give some statistical properties of the fragment population,
such as: number, total mass, mean mass, maximum mass, mean ratio between mass of
the most massive fragment and companion mass, average and maximum size, 
and maximum separation between the fragments. We find that the total mass in the fragments is in between 
$\sim 53$ \solm\ towards 16061--5048c1 and $\sim 4$\solm\ in 08477--4359c1,
in agreement with the significant amount of extended flux that has been resolved out, 
as shown in Table~\ref{tab_simba}. The average mass is of the order of the
mass of the Sun, and the most massive fragment is of $\sim 14$ \solm\ towards
16482--4443c2. The average sizes are around 0.01 -- 0.02 pc, corresponding
to $\sim 2000 - 4000$ au, while the fragmenting region is generally not very
compact, with maximum separation between the fragments of 0.05 -- 0.44 pc,
i.e.~$\sim 10000 - 90000$ au.

\begin{table*}[tbh]
\begin{center}
\caption[]{Some statistical properties of the fragment population in
each clump: number of fragments, total mass in fragments, average mass, maximum
mass, average ratio between mass of the most massive fragment 
and companion mass, average diameter, maximum diameter, and 
maximum separation between the intensity peaks.}
\label{tab_averages}
\small
\begin{tabular}{ccccccccc}
\hline \noalign {\smallskip}
Source  & fragment $n.$  & 
$m_{\rm tot}$ & $m_{\rm ave}$ & $m_{\rm max}$ & $< \frac{m_{\rm max}}{m_{\rm comp}}>$ &  $D_{\rm ave}$ & $D_{\rm max}$ & $S_{\rm max}$ \\
           &                                                                                             & \solm\      & \solm\     &  \solm\  &  &  pc    & pc & pc  \\
\hline \noalign {\smallskip}
08477--4359c1   & 4  & 3.7  &  0.9 &  1.5  &  3.1 & 0.013 &  0.02 & $\sim 0.10$   \\ 
15470--5419c1   & 14 & 24  &  1.7 &  12 & 34  & 0.018 &  0.03 & $\sim 0.44$  \\
15470--5419c3   & 9  &  31 &  3.4 &   10 &  23 & 0.026 &  0.04 & $\sim 0.34$  \\  
15557--5215c2   & 12  &  23 &  1.9 &  9.5 & 18 & 0.019 &  0.03 & $\sim 0.21$  \\
16061--5048c1   & 12 & 53 &  4.4 &  8.8 & 3.5 & 0.025 &  0.03 & $\sim 0.27$  \\
16061--5048c4   & 4   &  4.7  &  1.2 &  2.3 & 7.8 & 0.014 &  0.02 & $\sim 0.05$  \\ 
16482--4443c2   & 4   & 16 &  3.9 &  14 & 33 & 0.018 &  0.04 & $\sim 0.07$ \\
16573--4214c2   & 9   & 12 &  1.4 &   4.5 & 10  & 0.012 &  0.02 & $\sim 0.10$ \\
\hline \noalign {\smallskip}
\end{tabular}
\end{center}
\tablefoottext{a}{SFE not derived because the Herschel data are not available for this source.}
\end{table*}
\normalsize


\section{Discussion}
\label{discu}

\subsection{Fragment properties versus clump properties}
\label{discu_prop}

We have searched for possible relations between the properties of the
fragment population in Table~\ref{tab_averages}, and the physical parameters 
of the parent clumps (in Table~\ref{tab_simba}). We have focused on the following
clump parameters: gas temperature, total mass, diameter, H$_2$ total column 
density, CO depletion factor, non themal velocity dispersion, SFE, and
ratio between sound speed and non-thermal velocity dispersion. The non-thermal 
velocity dispersion, $\sigma_{\rm nth}$, has been estimated from the \CII\ (3--2) 
line width at half maximum by subtracting the thermal contribution (calculated 
assuming the gas temperature listed in Col.~7 of Table~\ref{tab_simba}). 
We stress from the beginning that all the conclusions drawn 
in this Section should be corroborated by a higher statistics. However,
some of our findings are indicative of possible correlations that will need
to be confirmed with statistically larger samples.

\begin{figure}[!]
\centerline{\includegraphics[width=9cm,angle=0]{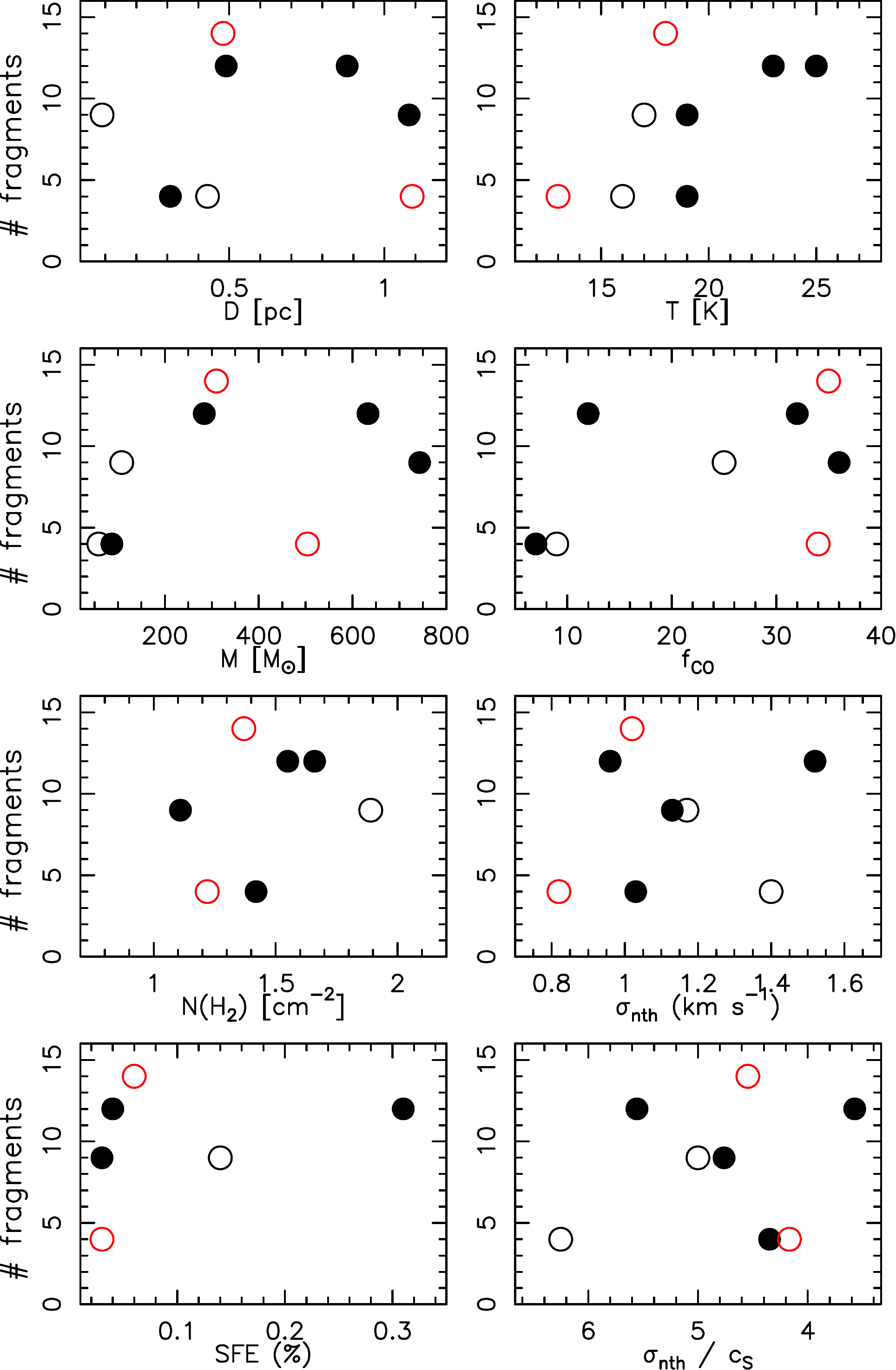}}
\caption{Number of fragments observed per clump (see Tables from \ref{tab_08477}
to \ref{tab_16573} versus the following clump parameters: diameter ($D$), mass
($M_{\rm gas}$), H$_2$ total column density (N(H$_2$)), SFE, gas temperature ($T$), 
CO depletion factor ($f_{\rm CO}$), non-thermal velocity dispersion ($\sigma_{\rm nth}$),
and the ratio between the non-thermal velocity
dispersion and the sound speed ($c_{\rm S}$), i.e. the Mach number.
Filled and empty circles indicate clumps with and without an embedded 24$\mu$m
source. The red circles corresponds to 15470--5419c1 and 16061--5048c4, in 
which the interpretation of the clump population needs to be taken with big caution 
(see Sect.~\ref{res_morf}). In the total H$_2$ column densities, we have excluded
the outlier 16482--4443c2 (see Table~\ref{tab_simba}).
}
\label{fig_corr_number}
\end{figure}

We have first investigated possible relations between the number of fragments
and the physical properties of the parent clump. This comparison is shown in 
Fig.~\ref{fig_corr_number}: overall, there are no clear (anti-)correlations, although
the sources with the largest temperature and mass tend to have more fragments.
In particular, with the exception of 16061--5048c4, clumps with more than 200 \solm\
always show at least 8 fragments. That the warmer clumps have, on average, more
fragments is consistent with the fact that the flux in the less massive fragments is
higher if they are warmer.
In Fig.~\ref{fig_corr_number}, we also distinguish
between the clumps with and without a 24~$\mu$m source, to check if
the presence of the embedded infrared source can influence the fragment
population. Again we cannot find any clear difference between the two classes of 
objects, which could indicate that the star formation activity does not influence the 
number of fragments (if we assume that the
presence of an embedded infrared source indicates a higher star formation activity). 
This finding is in agreement with Palau et al.~(\citeyear{palau}), whose
study suggested that the evolutionary stage does not have effects on the 
fragmentation.

\begin{figure}[!]
\centerline{\includegraphics[width=9cm,angle=0]{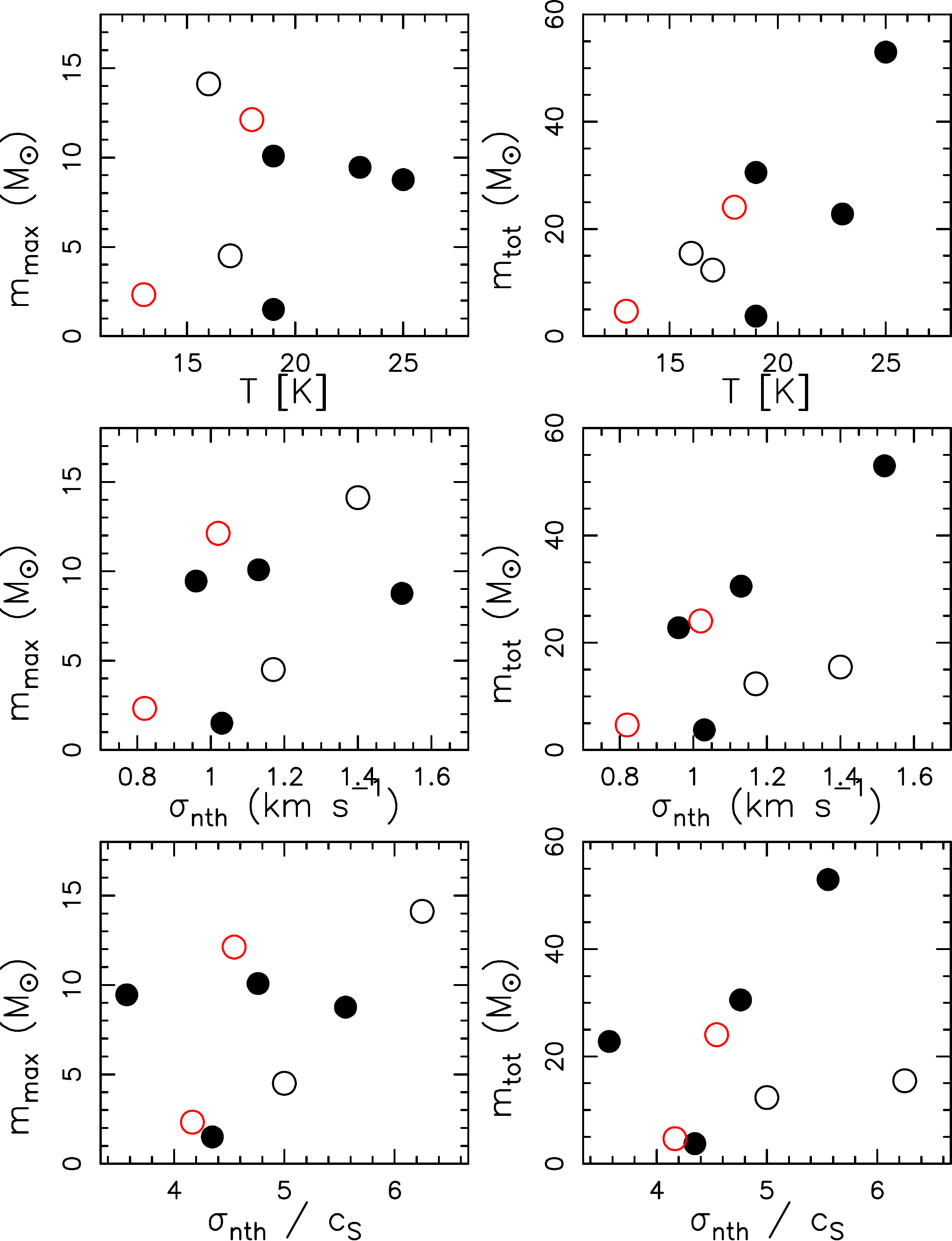}}
\caption{Maximum and total mass of the fragments ($m_{\rm max}$ and $m_{\rm tot}$,
respectively) as a function of the clump gas temperature ($T$), the non-thermal velocity dispersion
($\sigma_{\rm nth}$), and the ratio between $\sigma_{\rm nth}$
and the sound speed ($c_{\rm S}$), i.e. the Mach number.
The typical uncertainties on the masses, mainly due to the mass opacity coefficient, can be
up to a factor 2--3 (see Sect.~\ref{res_morf}).}
\label{masses_vs_temperature}
\end{figure}

We have investigated for possible relations between the mass of the fragments and
the clump properties. In Fig.~\ref{masses_vs_temperature}, we show the maximum
and total mass of the fragments ($m_{\rm max}$ and $m_{\rm tot}$, respectively)
as a function of $T$, $\sigma_{\rm nth}$, and the Mach number, 
i.e. the ratio between the non-thermal velocity dispersion and the sound speed, 
$\sigma_{\rm nth} / c_{\rm S}$, in order to evaluate the influence of the thermal
and turbulent supports. Both $m_{\rm max}$ and $m_{\rm tot}$ increase with $T$
and $\sigma_{\rm nth}$, suggesting that warmer and more turbulent clumps tend
to form more massive fragments. In order to evaluate which one is dominant, we
have plotted $m_{\rm max}$ and $m_{\rm tot}$ as a function of $\sigma_{\rm nth} / c_{\rm S}$:
even though in this case the trend is less apparent, the clumps with higher 
$\sigma_{\rm nth} / c_{\rm S}$, i.e. with lower thermal support, tend to form
more massive objects.

\begin{figure}[!]
\centerline{\includegraphics[width=9cm,angle=0]{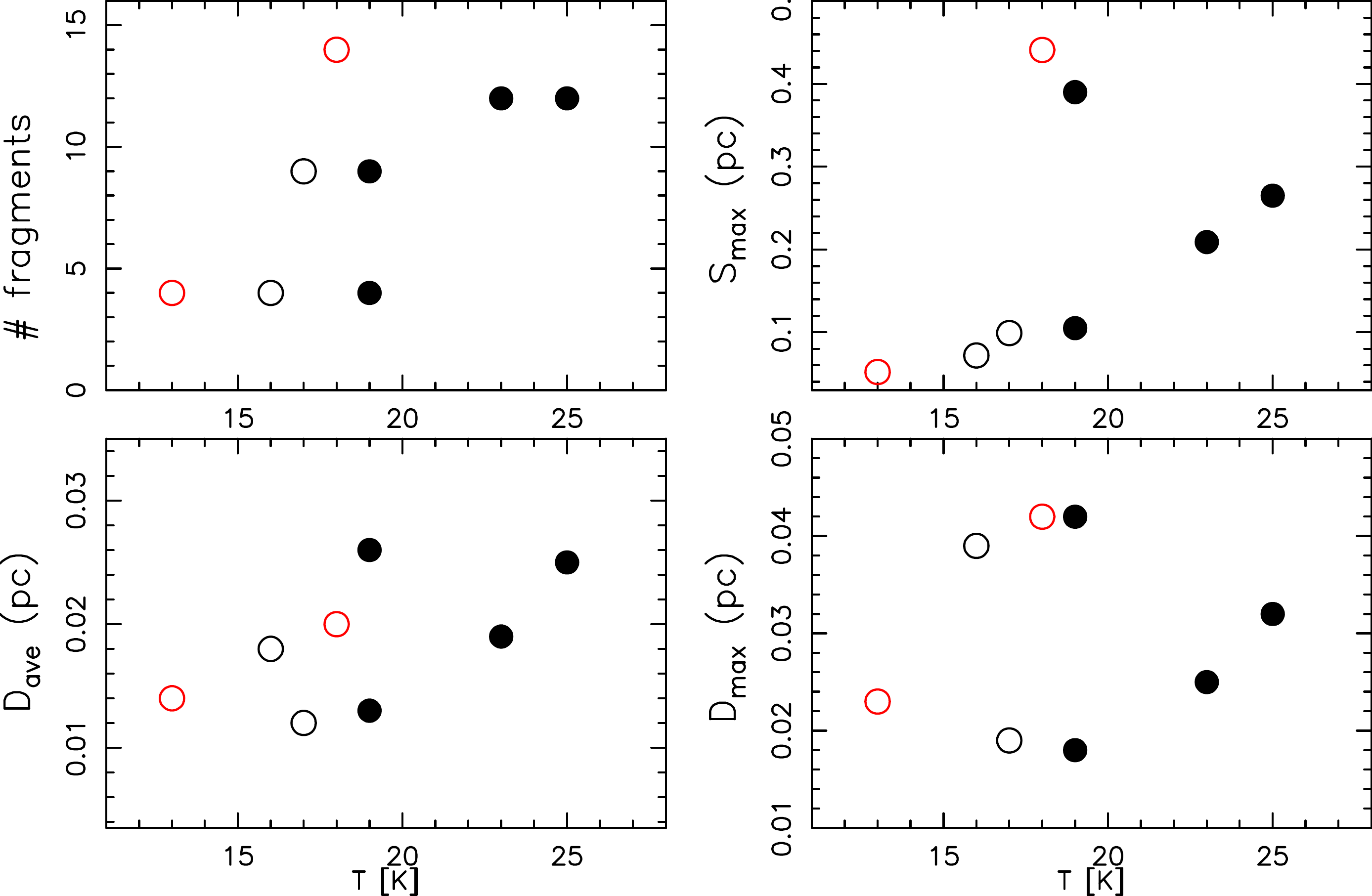}}
\caption{Number, maximum separation ($S_{\rm max}$),
average and maximum size of the fragments ($D_{\rm ave}$ and $D_{\rm max}$, respectively)
as a function of the clump temperature. The values on the y-axis are not rounded up to the 
significant digits, and the typical uncertainties on both the diameters and the separations are of
about 10$\%$.
}
\label{para_vs_temperature}
\end{figure}

The warmer clumps are also characterised by the largest separation between the fragments, 
as indicated by Fig.~\ref{para_vs_temperature}, in which we show the maximum linear separation 
($S_{\rm max}$) as a function of clump properties.
Finally, we have checked for possible trends with the average and maximum
clump linear size ($D_{\rm ave}$ and $D_{\rm max}$, respectively), and found again
a tentative positive trend with the clump temperature, although this result is quite
speculative and certainly needs to be corroborated by a higher statistics.


\subsection{Comparison with numerical simulations}
\label{discu_simu}

In Fontani et al.~(\citeyear{fontani2016}), we have simulated the gravitational
collapse of 16061--5048c1 through 3D numerical simulations adapted 
from Commer\c{c}on et al.~(\citeyear{commercon2011}) using the RAMSES code (Teyssier~\citeyear{teyssier2002}). 
We consider spherical clouds of radius $r_0$ with an initial density profile $\rho(r)=\rho_{\rm c}/(1+(r/r_{\rm c})^2)$, 
where $\rho_{\rm c}$ is the central density and $r_c$ the extent of the central plateau. In all models, we 
impose a density contrast of 10 between the center and the border of the cloud.
More details about the numerical model can be found in appendix~B.1 of Fontani et al.~(\citeyear{fontani2016}). 
The calculations were made adopting mass, temperature, average density, and turbulence of the parent
clump very similar to those measured in this source with single-dish observations
(Beltr\'an et al.~\citeyear{beltran2006}, Fontani et al.~\citeyear{fontani2012},
Giannetti et al.~\citeyear{giannetti}). We considered two degrees of magnetisation:
$\mu=2$, which is close to the values 2--3 that are observationally inferred (e.g.~Crutcher~\citeyear{crutcher}), 
and $\mu=200$, which corresponds to a quasi-hydrodynamical case. 
The outcome of the simulations were converted in flux density units of the
thermal dust continuum emission using the  RADMC-3D radiative transfer code (Dullemond et al.~\citeyear{dullemond2012}), 
following the same procedure as in Commer\c{c}on et al.~(\citeyear{commercon2012a},~\citeyear{commercon2012b}). 
These maps have then been post-processed through the CASA simulator to 
obtain synthetic images with the same observational conditions as the real 
observations assuming a source distance of 3.6~kpc and a region of 80 000$\times$80 000 au centred around the most massive protostar.
Following the same approach, in this paper we analyse a total of four reference models: 
\begin{itemize}
\item (1) initial mass of 100~\solm , gas temperature $T=10$~K, Mach
number $\mathcal{M}\sim 3$, and virial parameter $\alpha_{\rm vir}=2E_{\rm kin}/E_{\rm grav}=0.4$. The inital density profile is caracterized by $r_{\rm c}=0.22$~pc, $r_0=0.67$~pc, and $\rho_0=1.4\times10^{-20}$~g~cm$^{-3}$. The outcome of these simulations is shown in Fig.~\ref{fig_sim_100msun}. 
Note that these simulations correspond to the ones presented originally in Commer\c{c}on et al.~(\citeyear{commercon2011}),
which assume $\mu=130$ for the faint magnetised case. The difference with the $\mu=200$ case, assumed in the
other simulations, is completely irrelevant for the fragment population. 
They have been run without sink particles (e.g.~Bleuler \& Teyssier~\citeyear{bleuler}), 
thus without the protostellar radiative feedback;
\item (2) initial mass of 300~\solm , gas temperature $T=20$~K, Mach
number $\mathcal{M}\sim 6.4$, and virial parameter $\alpha_{\rm vir}=1.1$. The inital density profile is caracterized by $r_{\rm c}=0.085$~pc, $r_0=0.25$~pc, and $\rho_0=1.5\times10^{-18}$~g~cm$^{-3}$. The outcome of these simulations is shown in Fig.~\ref{fig_sim_300msun_T20_M6}. These simulations correspond to the ones used in  Fontani et al.~(\citeyear{fontani2016});
\item (3) initial mass of 300~\solm , gas temperature $T=20$~K, Mach
number $\mathcal{M}\sim 3$, and virial parameter $\alpha_{\rm vir}=0.22$. The initial density profile is the same as for simulations (2). The outcome of these simulations is shown in Fig.~\ref{fig_sim_300msun_T10_M3};
\item (4) initial mass of 300~\solm , gas temperature $T=10$~K, Mach
number $\mathcal{M}\sim 6.4$, and virial parameter $\alpha_{\rm vir}=1.1$. The inital density profile is caracterized by $r_{\rm c}=0.17$~pc, $r_0=0.5$~pc, and $\rho_0=1.9\times10^{-19}$~g~cm$^{-3}$. The outcome of these simulations is shown in Fig.~\ref{fig_sim_300msun_T10_M6}. In this set of initial conditions, we do not change the ratio between the initial thermal and gravitational energies of simulations (2) and (3). The initial clump is two times larger in radius and the initial density a factor of 8 smaller. Similarly, since the temperature is two times smaller and the Mach number does not change compared to simulations (2), the initial velocity fluctuations are a factor $\sqrt{2}$ smaller in amplitude. 
Simulations (2), (3), and (4) contain initially the same number of thermal Jeans masses ($\propto T^{3/2} \rho^{-1/2}$);
\end{itemize}

By comparing simulations (2), (3) and (4), we can understand the separate effect of 
temperature and turbulence. The source distance assumed in the synthetic images is 
always 3.6~kpc for the $M=100 - 300~$\solm\ cases, which is an average distance of 
the observed clumps. We have post-processed the simulations as made in
Fontani et al.~(\citeyear{fontani2016}).
A set of models that would reproduce the precise initial conditions of 
each single clump goes far beyond the scope of this paper. In the following, our main 
aim is to compare the overall morphology of the real and synthetic images
to understand if the observed population of fragments is more consistent with
strong or with faint magnetic support, and how the initial temperature and
turbulence induce clear differences in the population of the fragments.
To this purpose, we have decided to analyse the simulations stopped at
a SFE of $15\%$, which is an intermediate value between the minimum 
and maximum SFE found in our sample.

\begin{table*}[tbh]
\begin{center}
\caption[]{Statistical properties of the sink particles when the SFE is $15\%$: absolute time after the start of the simulations, time corresponding to an SFE of 15\% after the formation of the first sink particle (time $t_0$), number of sink particles, mean and maximum mass of the sink particles, mean and maximum separation between the sink particles, and SFR measured for an SFE $\in[4\%-15\%]$.}
\label{tab_simu_sink}
\small
\begin{tabular}{lccccccccc}
\hline \noalign {\smallskip}
Model  & $t_{15}$ & $t_{15}-t_0$ & $N_{\rm sink}$ &  $m_{\rm ave}$ & $m_{\rm max}$ &  $S_{\rm ave}$  & $S_{\rm max}$ & SFR \\
             & kyr  &  kyr                      &                            & \solm\                 &  \solm\                 &  au                        &  au & \solm~yr$^{-1}$ \\
\hline \noalign {\smallskip}
$\mu=2$, $T=20$K, $\mathcal{M}=6.4$, $\alpha_\mathrm{vir}=1.1$     &  108 &    38 & 36 & 1.6  & 7.6 & $1\times10^4$ & $2.6\times10^4$ & $1.5\times 10^{-3}$ \\ 
$\mu=2$, $T=10$K, $\mathcal{M}=6.4$, $\alpha_\mathrm{vir}=1.1$     & 302  &  106 & 47 & 1.3  & 6   & $2.2\times10^4$ & $5.1\times10^4$ &   $5.3\times 10^{-4} $\\
$\mu=2$, $T=20$K, $\mathcal{M}=3$, $\alpha_\mathrm{vir}=0.22$         & 98    &   28  & 44 & 1.7  & 18  & $4.5\times10^3$ & $1.3\times10^4$ &  $2.7\times 10^{-3} $\\
$\mu=200$, $T=20$K, $\mathcal{M}=6.4$, $\alpha_\mathrm{vir}=1.1$ &  84  &    35  & 71 & 0.9  & 3.3 & $1.1\times10^4$ & $3.3\times10^4$&  $1.7\times 10^{-3} $ \\
$\mu=200$, $T=10$K, $\mathcal{M}=6.4$, $\alpha_\mathrm{vir}=1.1$ & 237 & 107 & 84 & 0.65 & 3   & $2\times10^4$ & $6.2\times10^4$&  $6.2\times 10^{-4} $ \\
$\mu=200$, $T=20$K, $\mathcal{M}=3$, $\alpha_\mathrm{vir}=0.22$    & 78  &    27  & 47 & 1.4  & 6.6 & $2.3\times10^3$ & $1.5\times10^4$&  $2.5\times 10^{-3} $ \\
\hline \noalign {\smallskip}
\end{tabular}
\end{center}
\end{table*}
\normalsize

\subsubsection{Qualitative description of the simulations}
\label{qual_simu}

In this section, we briefly describe the outcome of the (2), (3), and (4) sets of simulations. The simulations (1) have been already analysed in Commer\c{c}on et al.~(\citeyear{commercon2011}) and we refer the readers to this work for more details. We focus on the sink particles (i.e., protostars) distribution properties. In the analysis, we select the sink particles with mass larger than $0.1$~\solm.
Table \ref{tab_simu_sink} reports the sink particles population properties for each simulation when the SFE is $\sim 15\%$. First we note that the time $t_{15}$ at which the SFE
reaches 15 \% after the start of the simulations depends on all the initial parameters: temperature, magnetisation, and Mach number. Nevertheless, if this time is rescaled after the time $t_0$ which corresponds to the time of the first sink particle formation, it does not depend on the magnetisation anymore. This result indicates that even though magnetic fields ``dilute'' gravity prior to the formation of the first protostars, the subsequent evolution of the SFE is mainly driven by the parent clump properties other than magnetic fields once gravity has taken over. Second, the mean and maximum mass are always largest in the strongly magnetised runs. Except for the $\mathcal{M}\sim 3$ runs, the number of sink particles is almost twice smaller with $\mu=2$, meaning that the strongly magnetised cases favour massive star formation, as already reported in the literature from both models and observations
(e.g., Commer\c{c}on et al.~\citeyear{commercon2011}, Tan et al.~\citeyear{tan2013}, Federrath et al.~\citeyear{federrath2015}, Kong et al.~\citeyear{kong2017},~\citeyear{kong2018}). The mean and maximum separations have been calculated within a spherical region of radius 40000~au around the most massive protostar in order to compare them with the observations. There is no evidence of a correlation of the separation, nor the mean mass with the initial parameters. We also report the measured star formation rate (SFR) which is computed for a SFE varying from 4\% to 15\%. By comparing simulations (2) and (3), the SFR decreases when $\alpha_\mathrm{vir}$ increases as expected from the analytical work in the literature (e.g., Hennebelle \& Chabrier~\citeyear{hennebelle2011b}). In the runs with $\mathcal{M} \sim 6.4$, the SFR increases by a factor $\sim 2.8$ as the initial temperature is doubled, in line with the total number of protostars. This factor is similar to the factor $\sqrt{8}$ resulting from the difference in the central density free-fall times of simulations (2) and (4) ($t_\mathrm{ff,(2)}=t_\mathrm{ff,(4)}/\sqrt{8}$), which implies that the SFR measured in unit of the freefall time, i.e., SFR$_\mathrm{ff}=\mathrm{SFR} t_\mathrm{ff}(\rho_{\rm c})/M_0$ (Krumholz \& McKee~\citeyear{krumholz2005}), remains unchanged.

\begin{figure*}[htp]
\includegraphics[width=9cm,angle=0]{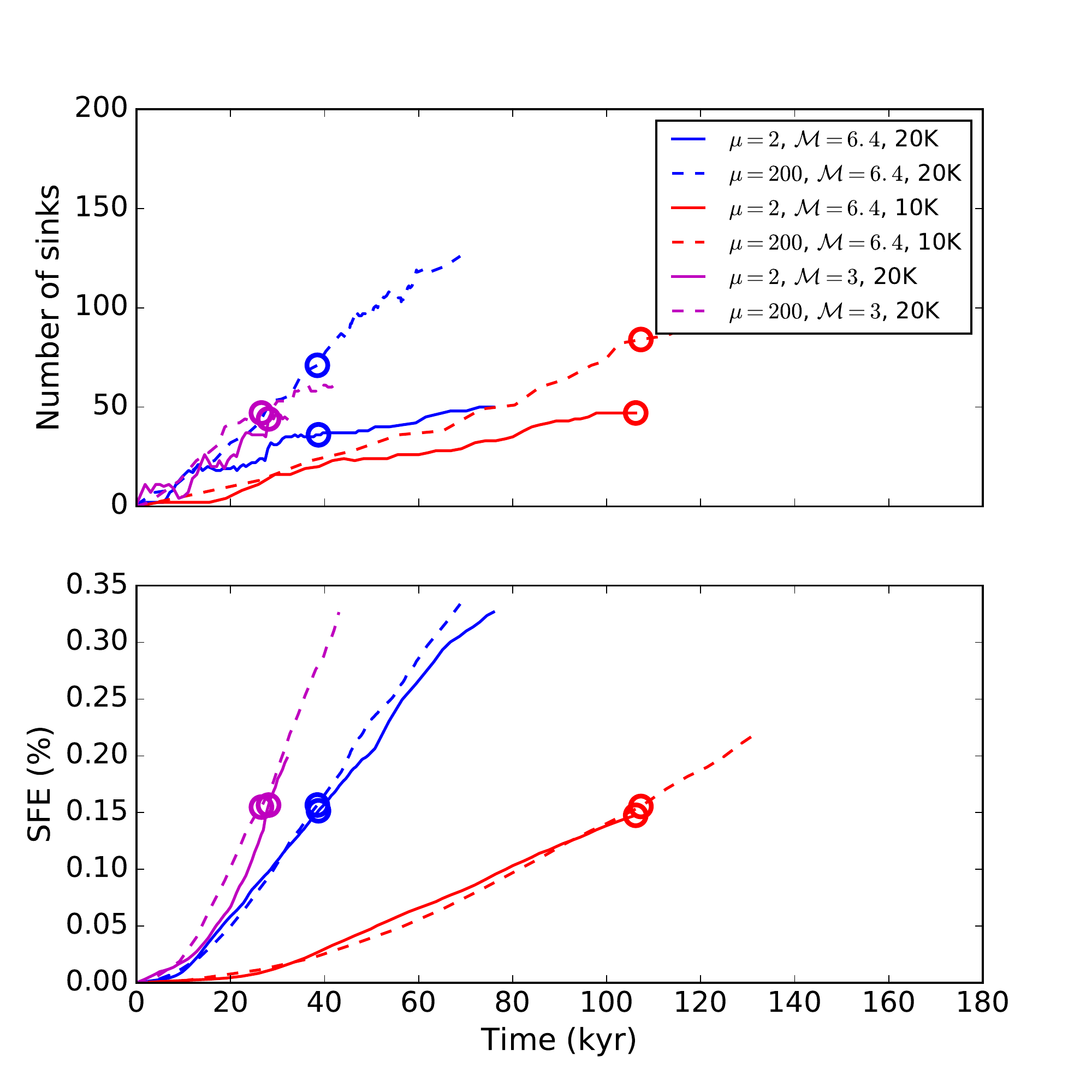}
\includegraphics[width=9cm,angle=0]{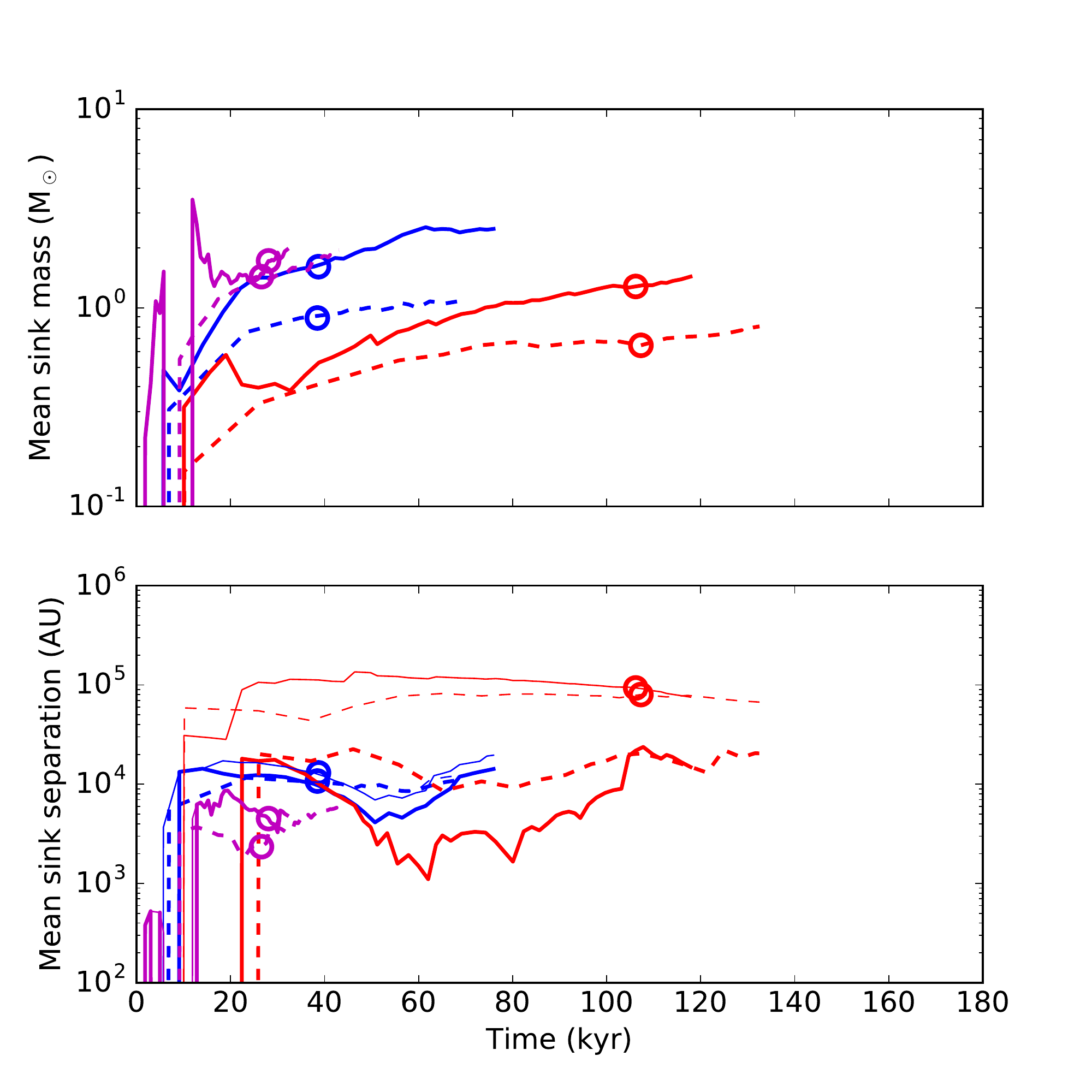}
\caption{{\it Left:} Time evolution of the number of sink particles (top) and of the SFE (bottom) for simulations (2), (3), and (4),
i.e. the simulations that reproduce the collapse of a 300~\solm\ clump. The circles indicate the time at which we postprocessed the simulations, which corresponds to a SFE$\simeq 15$\%. {\it Right:} Time evolution of the mean sink particle mass (top) and of the mean separation between sink particles (bottom). The thin lines show the mean separation calculated by accounting for the sink partciles located within a sphere of radius 40 000~au around the most massive one. The dashed lines show the mean separation between all the sink particles.}
\label{time:evol}
\end{figure*}

Figure~\ref{time:evol} (left) shows the time evolution of the number of sink particles and of the SFE for simulations (2), (3), and (4), in which the initial mass is 300~\solm. The time evolution has been rescaled after the time $t_0$ when the first sink particle was formed. First, we see that the number of protostars in the most turbulent runs ($\mathcal{M}\sim 6.4$) are the most sensitive to the magnetisation. We also note that runs with $\mu=2$ and $\mathcal{M}\sim 6.4$ exhibit a similar slope in the time evolution of the number of protostars. The number of protostars formed in the least turbulent runs $\mathcal{M}\sim 3$ is not dependent on the initial magnetisation. The initial turbulence being weaker, these runs are more dominated by gravity once collapse has been initiated. 
Interestingly, the SFE does not depend on the magnetisation, while it is  sensitive to the initial thermal and turbulent supports as previously mentioned.  

Figure~\ref{time:evol} (right) shows the time evolution of the mean sink mass and mean separation between sink particles. First, the mean and the maximum protostar mass are always larger in the strongly magnetised run in accordance with previous results of simulations (1). The evolution of the mean separation between protostars shows interesting features. If we focus on the solid lines, i.e., all the sink particles which sit within a sphere of radius 40000~au around the most massive one, we do not find a clear correlation with the initial parameters. The mean separation of the  $\mathcal{M}\sim 6.4$ runs is globally smaller for the strongly magnetised case. The dashed line represents the evolution of the mean separation if all the sink particles formed in the simulations are considered. The separation then depends more on the initial temperature and Mach number than on the magnetisation. Focusing on $\mathcal{M}\sim 6.4$ runs, the separation is a factor $\sim 10$ larger in the case of a lower initial temperature, with a maximum separation larger than $80 000$ au. This result is consistent with previous studies that show how the fragmentation region extent depends strongly on the initial density profile (e.g., Girichidis et al.~\citeyear{girichidis}). In simulations (4), the inital density profile is flatter than in simulations (2) which favors fragmentation over a wider region. This means that some parts of the simulated clumps where star formation takes place are not taken into account in the synthetic observation we present below. Last but not least, the analysis on the mean separation, averaged over all dimensions, does not reflect the morphology of the fragmentation regions. In Appendix~B, we show histrograms of the sink particle separation distribution, as well as their 2D projected distributions for the  $\mathcal{M}\sim 6.4$, $T=10$~K runs. A discussion of these distributions is also provided in the same Appendix.

\subsubsection{Qualitative comparison with observations}
\label{qual_obs}

\begin{figure}[!]
\centerline{\includegraphics[width=9cm,angle=0]{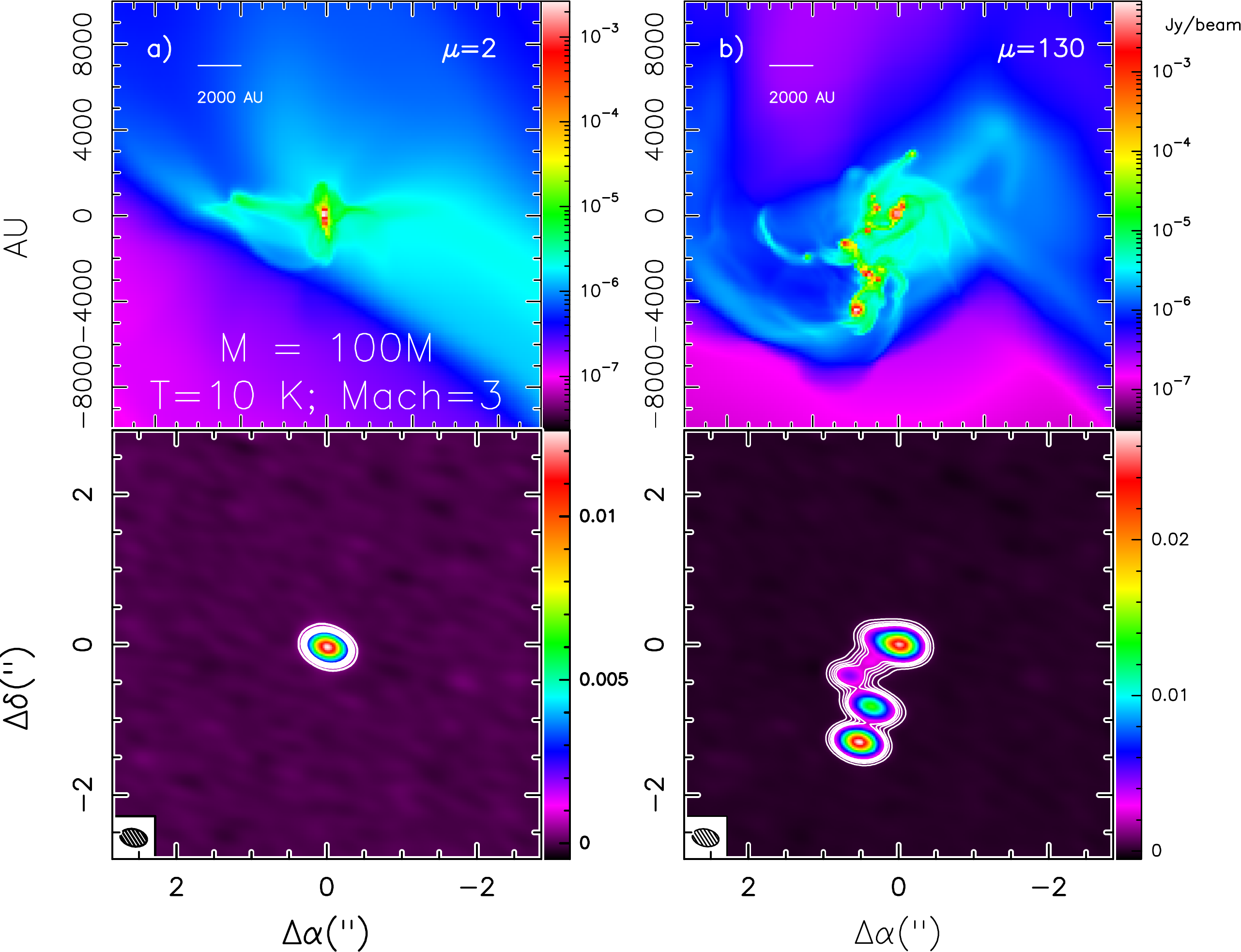}}
\caption{Simulations of the gravitational collapse of a 100 \solm\ collapsing clump, with
gas temperature $T=10$~K, and Mach number $\mathcal{M}\sim 3$. 
{\bf a)}: model predictions for the strongly magnetised case, $\mu=2$. In the top 
panel, we show the direct outcome of the simulations, while the same image 
processed with the CASA simulator is shown in the bottom panel (first contour and
step is 0.3 mJy beam$^{-1}$).
{\bf b)}: same as {\bf a)} for the weakly magnetised case, $\mu=130$.
}
\label{fig_sim_100msun}
\end{figure}

We now compare our observations with the synthetic images created from the simulations
described in the previous Section with the method explained in Fontani et al.~(\citeyear{fontani2016}). 
Let us discuss first the case in Fig.~\ref{fig_sim_100msun}: assuming a clump mass
of 100 \solm , the model is the most appropriate to reproduce the initial conditions 
of the less massive clumps of our sample, i.e. 08477--4359c1,16482--4443c2, and 16573--4214c2. 
Depending on $\mu$, the simulations predict 
either one single fragment in the high magnetic-support case ($\mu=2$), or several 
fragments packed in a region smaller than $\sim 8000$ au in the other case ($\mu=130$,
see bottom panels in Fig.~\ref{fig_sim_100msun}).
Both predictions are different from our images, because the three clumps mentioned above
show all more than one fragment, but these are distributed in an area more extended 
than 8000 au ($\sim 15000 - 20000$ au, see Fig.~\ref{fig_maps_tot} and Col.~8 in 
Table~\ref{tab_averages}). However, the case that better resembles 
the images of the less massive sources is the strongly magnetised case, $\mu=2$,
because in the $\mu=130$ case, the fragments should have
similar size and flux, while in our objects all clumps have a dominant fragment surrounded
by much fainter fragments. Moreover, our simulations assume, among the initial conditions, that 
a single, spherically symmetric clump fragments. Models assuming more complex density profiles 
such as, e.g., turbulent periodic boxes (e.g.~Padoan \& Nordlund~\citeyear{padoan}, Federrath \& 
Klessen~\citeyear{federrath2012}, Haugb{\o}lle et al.~\citeyear{haugbolle}, Mocz et al.~\citeyear{mocz}), 
or clouds with more complex turbulent structure (e.g.~Li et al.~\citeyear{li}, Girichidis
et al.~\citeyear{girichidis}, Federrath et al.~\citeyear{federrath2014}, Myers et al.~\citeyear{myers2014}), 
would provide certainly more fragments. Indeed, our targets could not be single spherical objects,
as can be deduced also from the SIMBA maps in Fig.~\ref{fig_simba}.
Hence, the initial conditions in our simulations are expected to be those that show 
the lower level of fragmentation, and the comparison needs to be taken with caution. 

\begin{figure}[!]
\centerline{\includegraphics[width=9cm,angle=0]{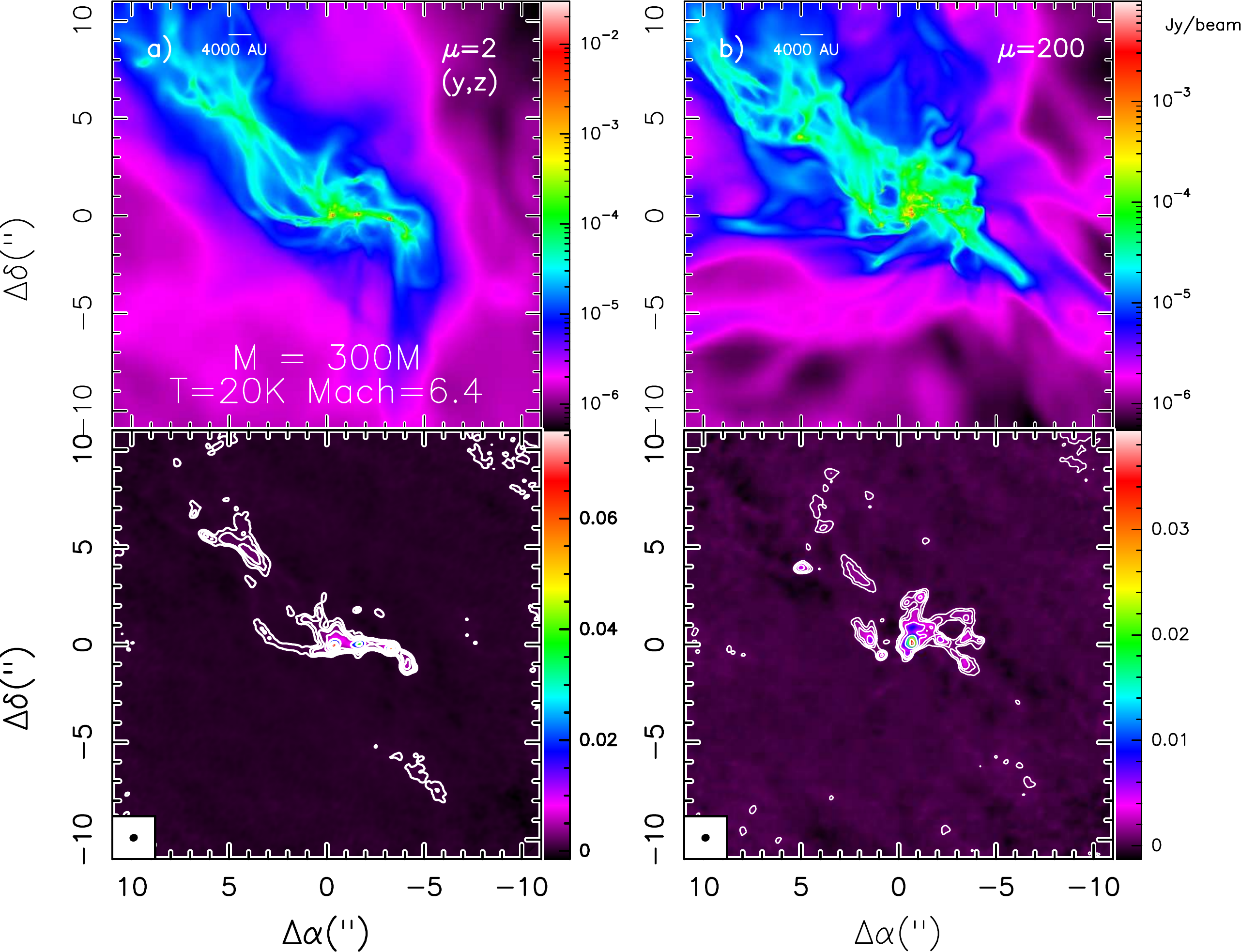}}
\caption{Simulations of the gravitational collapse of a 300 \solm\ collapsing clump, 
with gas temperature $T=20$~K, and Mach number $\mathcal{M}\sim 6.4$. 
{\bf a)}: model predictions for the strongly magnetised case , $\mu=2$. In the top 
panel, we show the direct outcome of the simulations, while the same image 
processed with the CASA simulator is shown in the bottom panel. The first contour is
0.6 mJy beam$^{-1}$ as in Fontani et al.~(\citeyear{fontani2016}, roughly
3 times the 1$\sigma$ rms of most images). The steps are 1.2, 2, 5, 10, 30 and
50 mJy beam$^{-1}$.
{\bf b)}: same as {\bf a)} for the weakly magnetised case, $\mu=130$.
These simulations have the same set of initial parameters as those discussed in Fontani et 
al.~(\citeyear{fontani2016}), but we have post-processed and analysed those at which
the star formation efficiency (SFE) is $15\%$, while in Fontani et al.~(\citeyear{fontani2016})
we analysed the time at which the SFE matched that of IRAS 16061--5048c1
($\sim 30\%$).
}
\label{fig_sim_300msun_T20_M6}
\end{figure}

The case shown in Fig.~\ref{fig_sim_300msun_T20_M6}, especially made to match
as well as possible the parameters of 16061--5048c1 in Fontani et al.~(\citeyear{fontani2016}),
can be adopted to qualitatively discuss also 15470--5419c1, 15470--5419c3, and 15557--5215c2. 
The only difference with Fontani et al.~(\citeyear{fontani2016})
is that the image that we analyse in this work is obtained when the SFE is $15\%$,
while that analysed in Fontani et al.~(\citeyear{fontani2016}) matched the total
flux observed towards 16061--5048c1.
For 16061--5048c1, we concluded that the overall filamentary morphology was a
strong evidence in favour of the $\mu=2$ case, which cannot be obtained in a
weakly magnetised case (Fontani et al.~\citeyear{fontani2016}). A filamentary-like shape
is found also in 15470--5419c3. The other two sources (15470--5419c1 and 15557--5215c2) 
show a more irregular structure, which could be explained by a weakly magnetised clump.
But even in this case there is not a good agreement, because the fragments predicted by 
the simulations are distributed in an area less extended than that found in our ALMA images. 
Moreover, the case of 15470--5419c1 must be interpreted with particular caution because 
of the huge amount of extended flux resolved out and the location of the most massive
fragments at the border of the primary beam.

\begin{figure}[!]
\centerline{\includegraphics[width=9cm,angle=0]{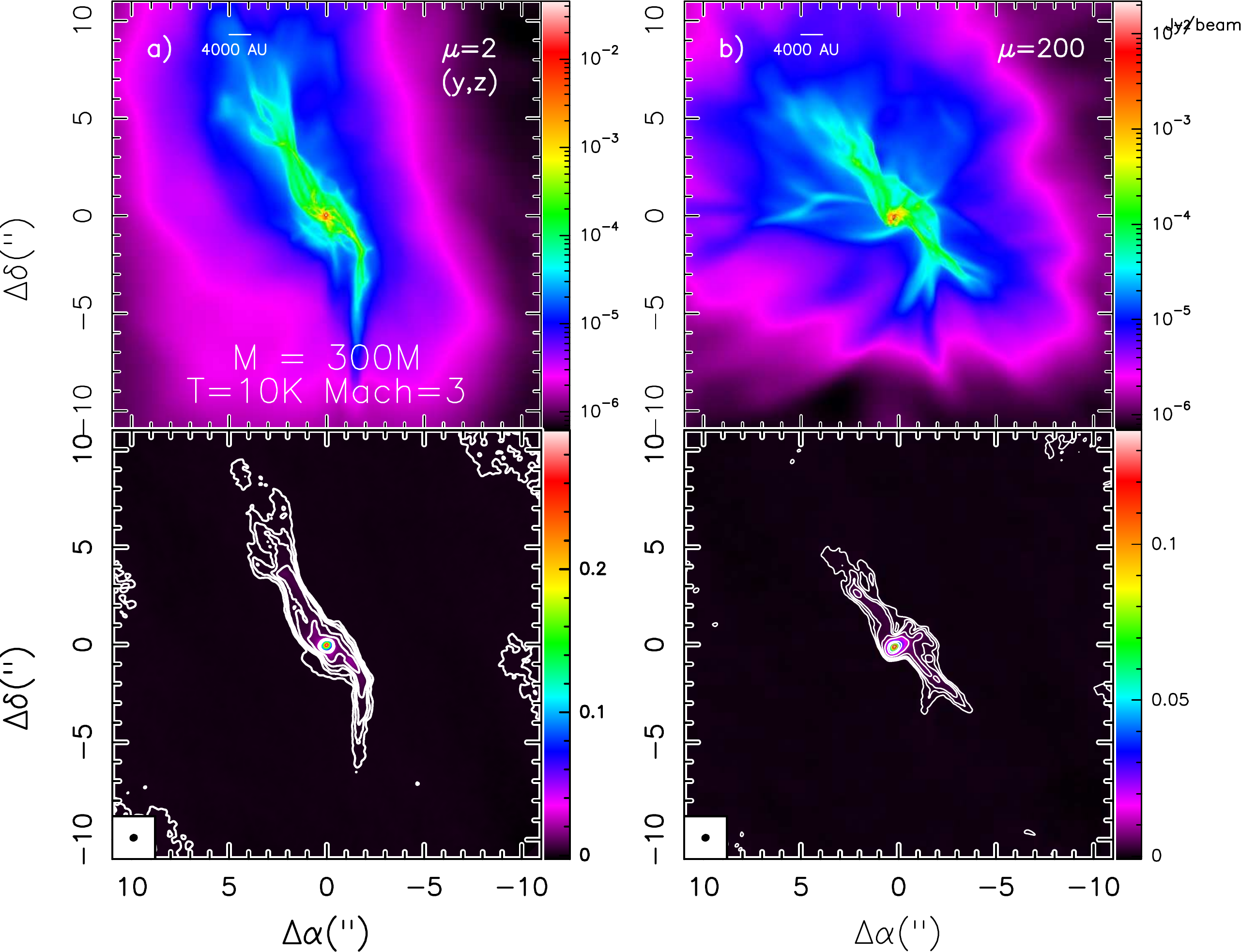}}
\caption{Same as Fig.~\ref{fig_sim_300msun_T20_M6} for a clump with
Mach number $\mathcal{M}\sim 3$. }
\label{fig_sim_300msun_T10_M3}
\end{figure}

\begin{figure}[!]
\centerline{\includegraphics[width=9cm,angle=0]{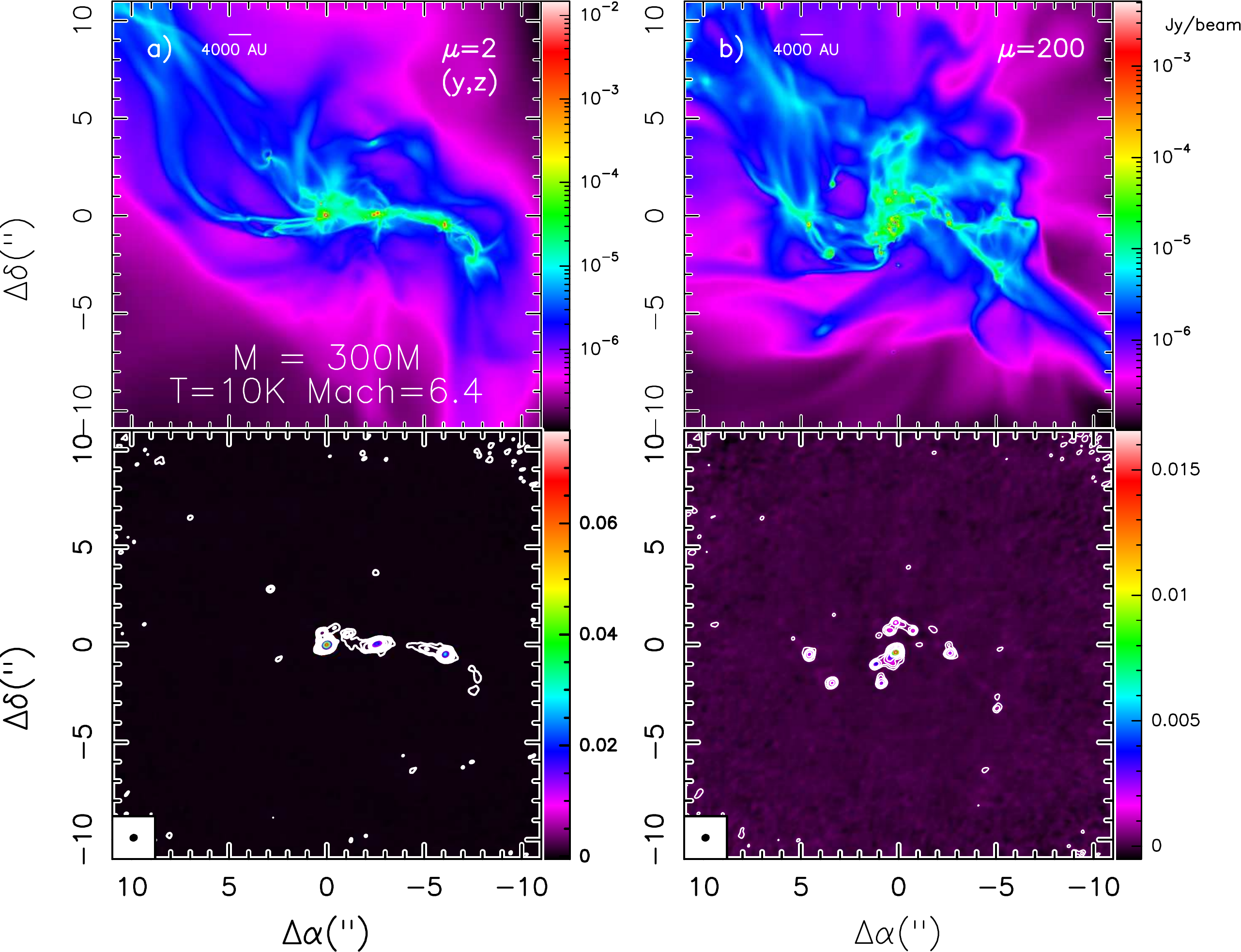}}
\caption{Same as Fig.~\ref{fig_sim_300msun_T20_M6} for a clump with
gas temperature $T=10$~K. 
}
\label{fig_sim_300msun_T10_M6}
\end{figure}

Figs.~\ref{fig_sim_300msun_T10_M3} and \ref{fig_sim_300msun_T10_M6} show
what happens when we start from a lower Mach number and a lower kinetic temperature,
respectively. Inspection of these figures indicates that more than one fragment 
can be found only if the turbulence is relatively high, because in the $\mathcal{M}=3$ 
case we find no fragmentation, 
independent of the magnetic field strength.
None of our sources, however, show less than 4 fragments,
which implies that this combination of initial conditions is not realistic. This is
consistent with the clump velocity dispersions, which indicate always high levels of
turbulence. However, as discussed above for the 100 \solm\ case,
a big caveat arises from the initial density profile adopted
in the simulations, which, as stated before, is expected to provide the lower
number of fragments and could not be appropriate for our sources if
they are not single global spherically symmetric clumps.

To make a more quantitative comparison with the data, we have calculated the properties
of the fragments using the same criteria adopted for the real images (see Sect.~\ref{res_prop}).
Some statistically relevant quantities are reported in Table~\ref{tab_simu_averages}, 
and confirm the previous qualitative analysis, namely that: (1) the $\mu=2$ case
produces fewer and more massive fragments; (2) more than 1 fragment is possible
only if the turbulence is higher ($\mathcal{M}\sim 6.4$ case); (3) the initial temperature has 
limited influence on the final population of fragments, but warmer clumps tend to exhibit
more fragments because the fragmentation region is more concentrated. As shown in figures \ref{time:evol} and \ref{fig:sep_proj}, some part of the fragmentation region is missed by our analysis of the synthetic maps of simulations (4) if we consider only the region that would have been observed with ALMA. Overall, the synthetic images discussed in this work allow us to confirm 
that both the turbulence and the magnetic field are key ingredients in the fragmentation 
of massive dense clumps, and our observations tend to favour an interplay between
turbulence and magnetic field to explain both the morphology and the number of
fragments detected.

\begin{table*}[tbh]
\begin{center}
\caption[]{Statistical properties of the fragment population in the synthetic
images. The same properties as in Table~\ref{tab_averages} are shown.
The SFE assumed in each simulation is $15\%$.}
\label{tab_simu_averages}
\small
\begin{tabular}{lcccccccc}
\hline \noalign {\smallskip}
Model  & fragment $n.$  & $m_{\rm tot}$ & $m_{\rm ave}$ & $m_{\rm max}$ & $< \frac{m_{\rm max}}{m_{\rm comp}}>$ &  $D_{\rm ave}$ & $D_{\rm max}$ & $S_{\rm max}$ \\
           &                                & \solm\      & \solm\               &  \solm\              &                          &  pc    & pc & pc - au  \\
\hline \noalign {\smallskip}
$\mu=2$, $T=20$K, $\mathcal{M}=6.4$ & 9  & 62 & 7  & 26 & 35  & 0.023 & 0.048 & 0.21 - $\sim 41000$ \\
$\mu=2$, $T=10$K, $\mathcal{M}=6.4$ & 6  & 75 & 12 & 38 & 38 & 0.018 & 0.026 & 0.17 - $\sim 33000$ \\
$\mu=2$, $T=20$K, $\mathcal{M}=3$ & 1  & 173 & -- & -- & -- & 0.09 & -- & -- \\
$\mu=200$, $T=20$K, $\mathcal{M}=6.4$ & 13  & 45 & 3.5 & 25 & 24 & 0.020 & 0.041 & 0.21 - $\sim 41000$ \\
$\mu=200$, $T=10$K, $\mathcal{M}=6.4$ & 11 & 25 & 12 & 13 & 22 & 0.013 & 0.022 & 0.18 - $\sim 35000$ \\
$\mu=200$, $T=20$K, $\mathcal{M}=3$ & 1 & 87 & -- & --  & -- & 0.083 & -- & -- \\
\hline \noalign {\smallskip}
\end{tabular}
\end{center}
\end{table*}
\normalsize


\section{Conclusions}
\label{conc}

We have used ALMA to image the 278~GHz continuum emission in 11 massive dense clumps in which 
the star formation activity is low or absent, to understand the fragment population at the earliest
phases of the gravitational collapse. The angular resolution of our observations ($0\farcs 25$)
is able to resolve a linear scale of $\sim 1000$ au at the distance of the sources.
The clumps show a fragment population with at least four fragments distributed in different morphologies, 
mostly filament-like or irregular.  In four targets a dominant fragment surrounded by companions 
with much smaller mass and size is identified, while many ($\geq 8$) fragments with a gradual 
change in masses and sizes are found in the others. The number of fragments is likely a lower 
limit given the huge amount of missing flux in most of the sources. This effect is especially
relevant in the targets showing a displacement between the phase centre and 
the location of the ATLASGAL emission peak. In general, there are no clear relations between 
the properties of the clumps and those of their fragments, although our results tentatively indicate 
that the more massive and warmer clumps tend to have more fragments concentrated over a single region.
Comparison with the simulations indicate that fragmentation of clumps with
initial conditions similar to our objects can occur only assuming a high 
($\mathcal{M}\sim 6$) initial turbulence, while in a lower turbulent scenario ($\mathcal{M}\sim 3$)
only one very massive fragment surrounded by an extended envelope is expected.
Both observations and simulations show that the initially warmer clumps tend to form more fragments. 
A filament-like morphology is predicted most likely in a highly magnetised clump. We 
hence conclude that the clumps with many fragments distributed in a filamentary-like 
structure can be obtained only if the magnetic field plays a dominant role, while the 
other morphologies are possible also in a weaker magnetised case, or in a scenario
in which both magnetic field and turbulence interplay.

{\it Acknowledgments.} 
This paper makes use of the following ALMA data: ADS/JAO.ALMA.2012.1.00366.S. 
ALMA is a partnership of ESO (representing its member states), NSF (USA) and NINS 
(Japan), together with NRC (Canada), NSC and ASIAA (Taiwan), and KASI (Republic of Korea), 
in cooperation with the Republic of Chile. The Joint ALMA Observatory is operated by ESO, 
AUI/NRAO and NAOJ. We acknowledge the Italian-ARC node for their help in the reduction 
of the data. We acknowledge partial support from Italian Ministero dell'Istruzione, Universit\'a 
e Ricerca through the grant Progetti Premiali 2012 $-$ iALMA (CUP C52I13000140001) 
and from Gothenburg Centre of Advanced Studies in Science and Technology through 
the program {\it Origins of habitable planets}. ASM is partially supported by the Deutsche 
Forschungsgemeinschaft (DFG) through grant SFB956 (subproject A6).

{}

\clearpage

\renewcommand{\thetable}{A-\arabic{table}}
\renewcommand{\thefigure}{A-\arabic{figure}}
\renewcommand{\thesection}{A-\arabic{section}}
\setcounter{table}{0}
\setcounter{figure}{0}
\setcounter{section}{0}
\section*{Appendix A: Identification and physical properties of the fragments}
\label{appA}

\begin{figure}[!]
\centerline{\includegraphics[width=9cm,angle=0]{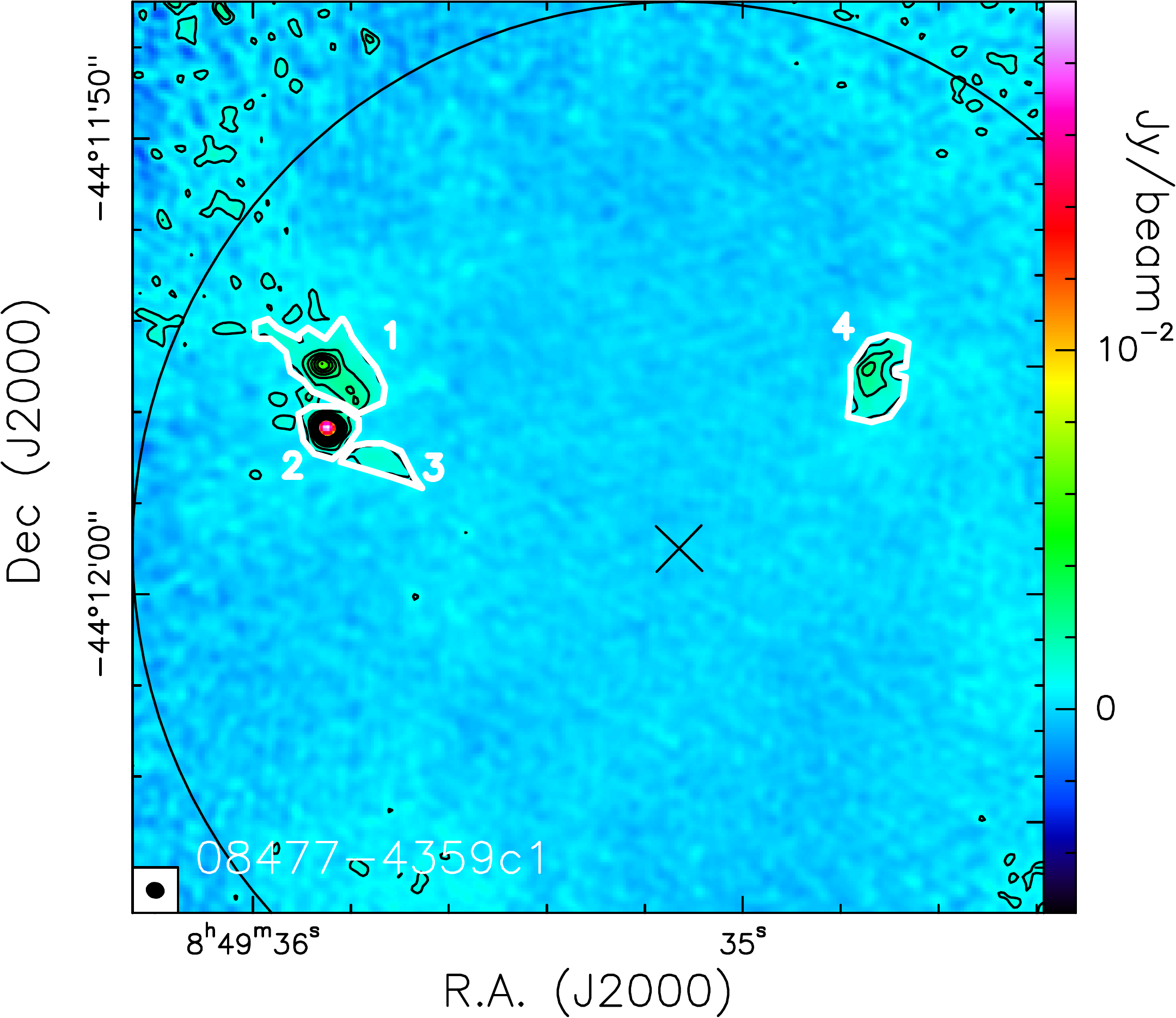}}
\caption{ALMA dust thermal continuum emission map at 278~GHz towards 08477--4359c1. 
The first contour level, and the step, is $8.7\times 10^{-4}$ Jy beam$^{-1}$, corresponding
to the 3$\sigma$ rms noise level ($1\sigma \sim 2.9\times 10^{-4}$ Jy beam$^{-1}$).
The white polygons indicate the fragments identified on the basis of the criteria described 
in Sect.~\ref{res}.
In each panel, the circle indicates the ALMA field of view at 278~GHz ($\sim$24\asec ),
and the cross the phase center, corresponding to the coordinates in Table~\ref{tab_simba}.
}
\label{fig_map_08477}
\end{figure}

\begin{figure}[!]
\centerline{\includegraphics[width=9cm,angle=0]{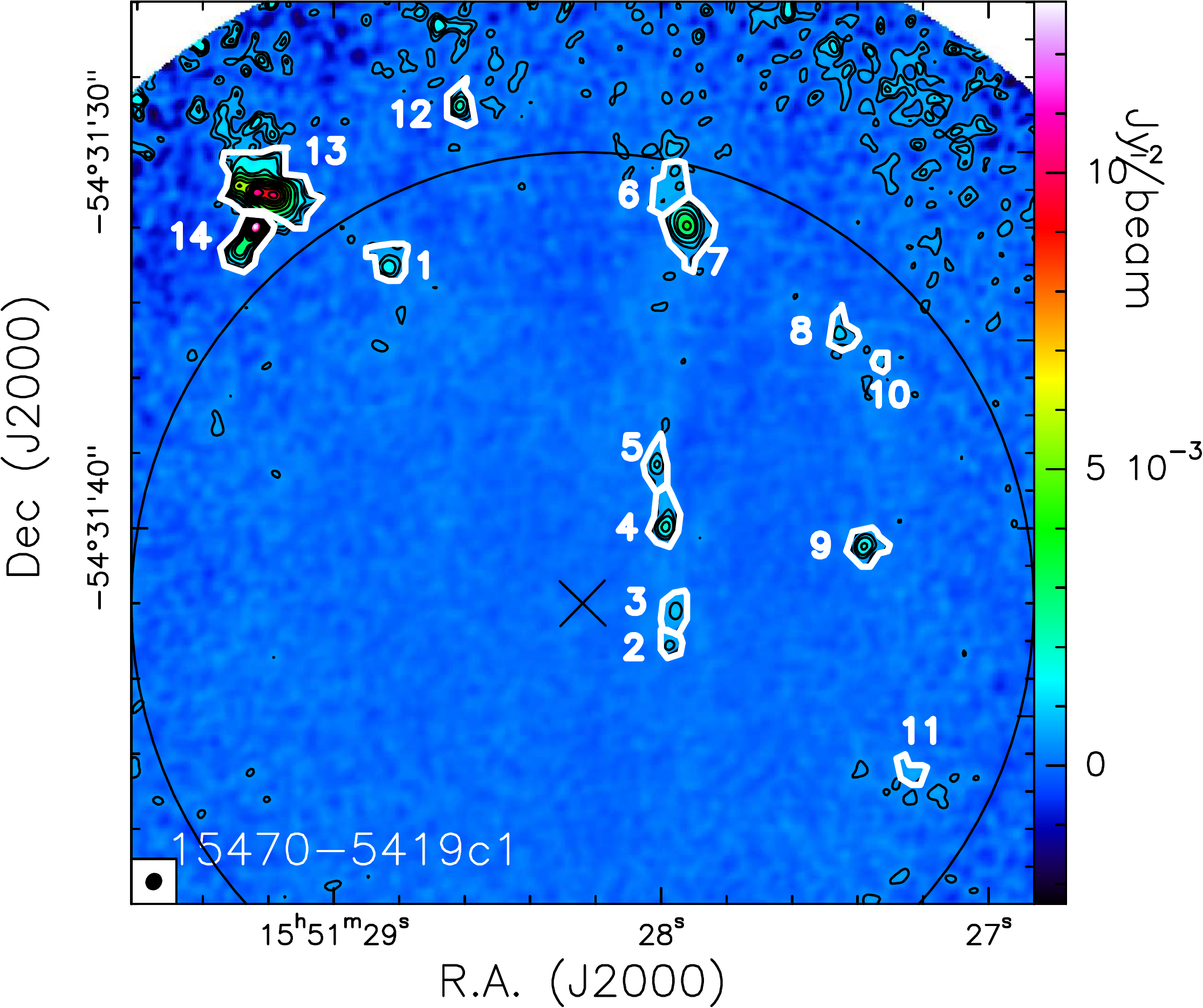}}
\caption{Same as Fig.~\ref{fig_map_08477} for 15470--5419c1. 
The first contour level, and the step, is $3.6\times 10^{-4}$ Jy beam$^{-1}$, corresponding
to the 3$\sigma$ rms noise level ($1\sigma \sim 1.2\times 10^{-4}$ Jy beam$^{-1}$).
}
\label{fig_map_15470c1}
\end{figure}

\begin{figure}[!]
\centerline{\includegraphics[width=9cm,angle=0]{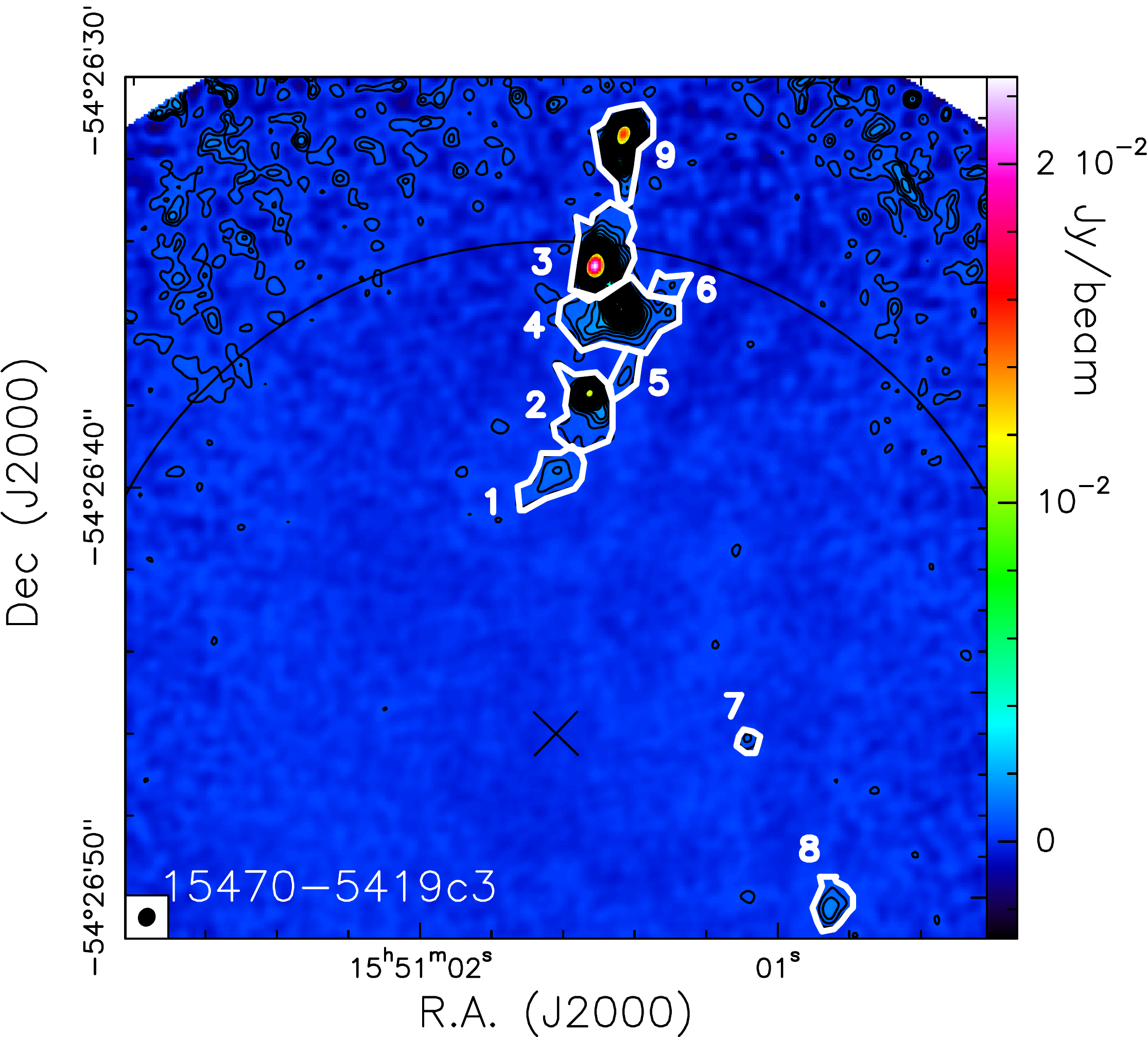}}
\caption{Same as Fig.~\ref{fig_map_08477} for 15470--5419c3. 
The first contour level, and the step, is $4.2\times 10^{-4}$ Jy beam$^{-1}$, corresponding
to the 3$\sigma$ rms noise level ($1\sigma \sim 1.4\times 10^{-4}$ Jy beam$^{-1}$).
}
\label{fig_map_15470c3}
\end{figure}

\begin{figure}[!]
\centerline{\includegraphics[width=9cm,angle=0]{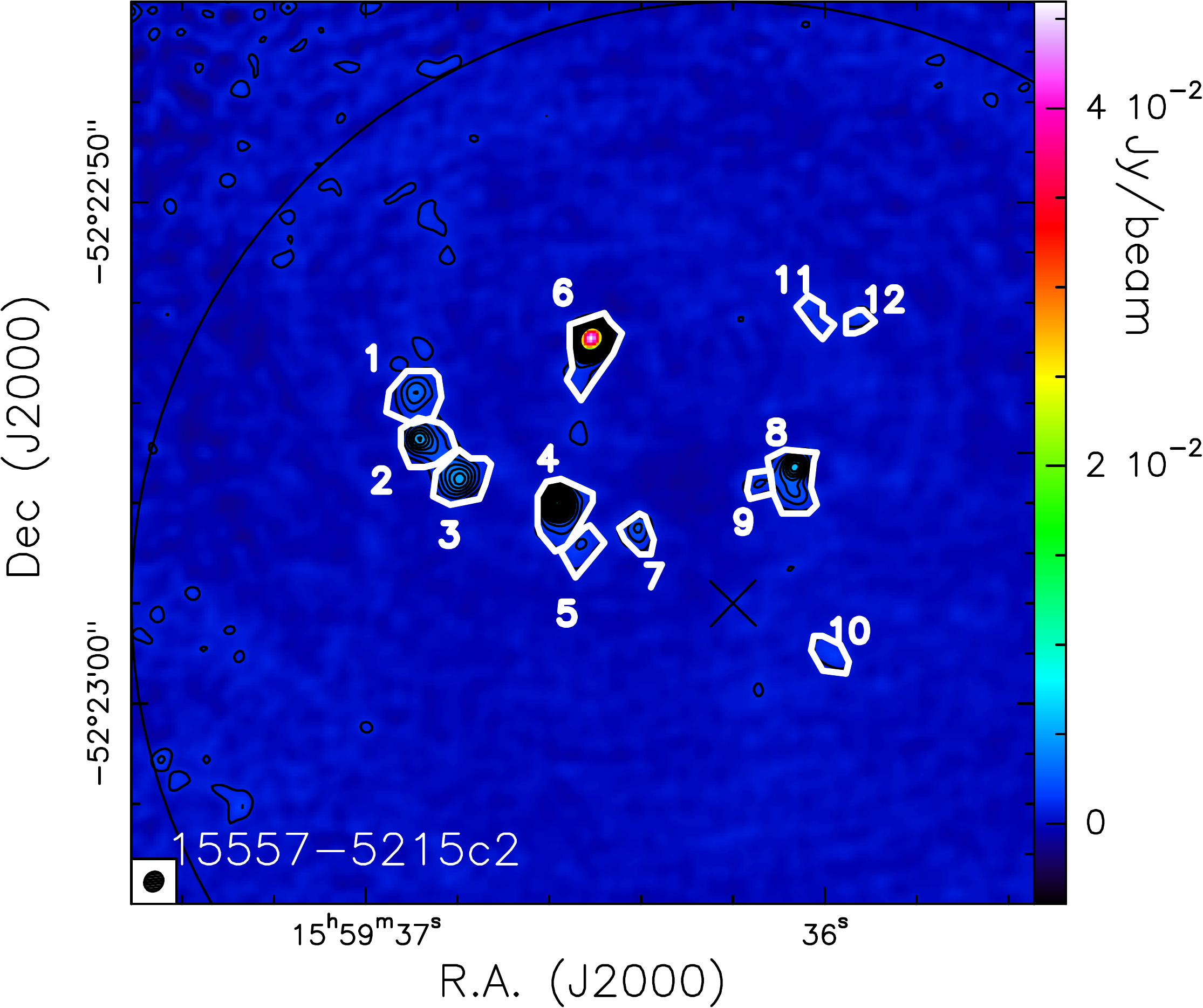}}
\caption{Same as Fig.~\ref{fig_map_08477} for 15557--5215c2. 
The first contour level, and the step, is $7.8\times 10^{-4}$ Jy beam$^{-1}$, corresponding
to the 3$\sigma$ rms noise level ($1\sigma \sim 2.6\times 10^{-4}$ Jy beam$^{-1}$).
}
\label{fig_map_15557c2}
\end{figure}

\begin{figure}[!]
\centerline{\includegraphics[width=9cm,angle=0]{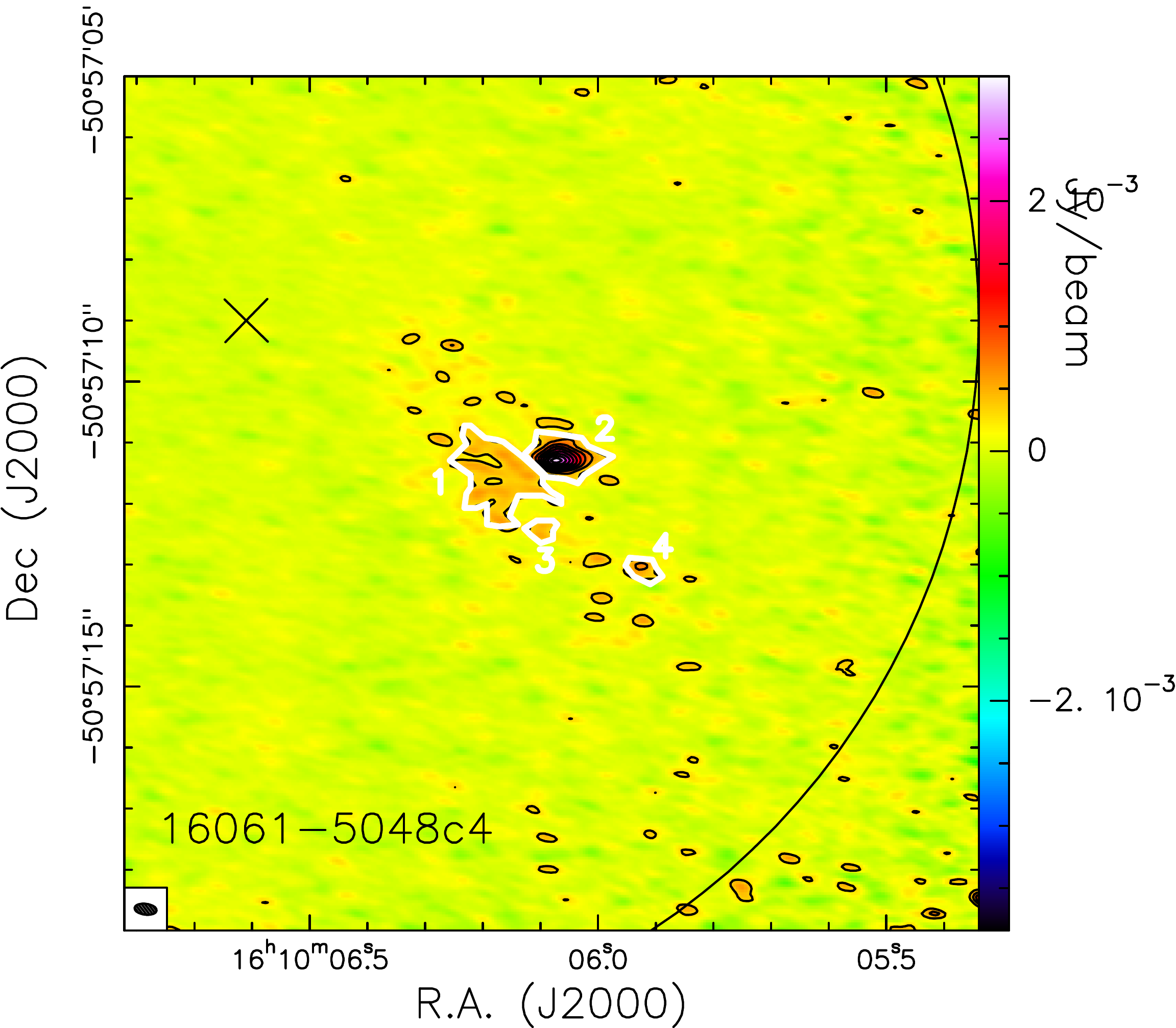}}
\caption{Same as Fig.~\ref{fig_map_08477} for 16061--5048c4. 
The first contour level, and the step, is $3\times 10^{-4}$ Jy beam$^{-1}$, corresponding
to the 3$\sigma$ rms noise level ($1\sigma \sim 1\times 10^{-4}$ Jy beam$^{-1}$).
}
\label{fig_map_16061}
\end{figure}

\begin{figure}[!]
\centerline{\includegraphics[width=9cm,angle=0]{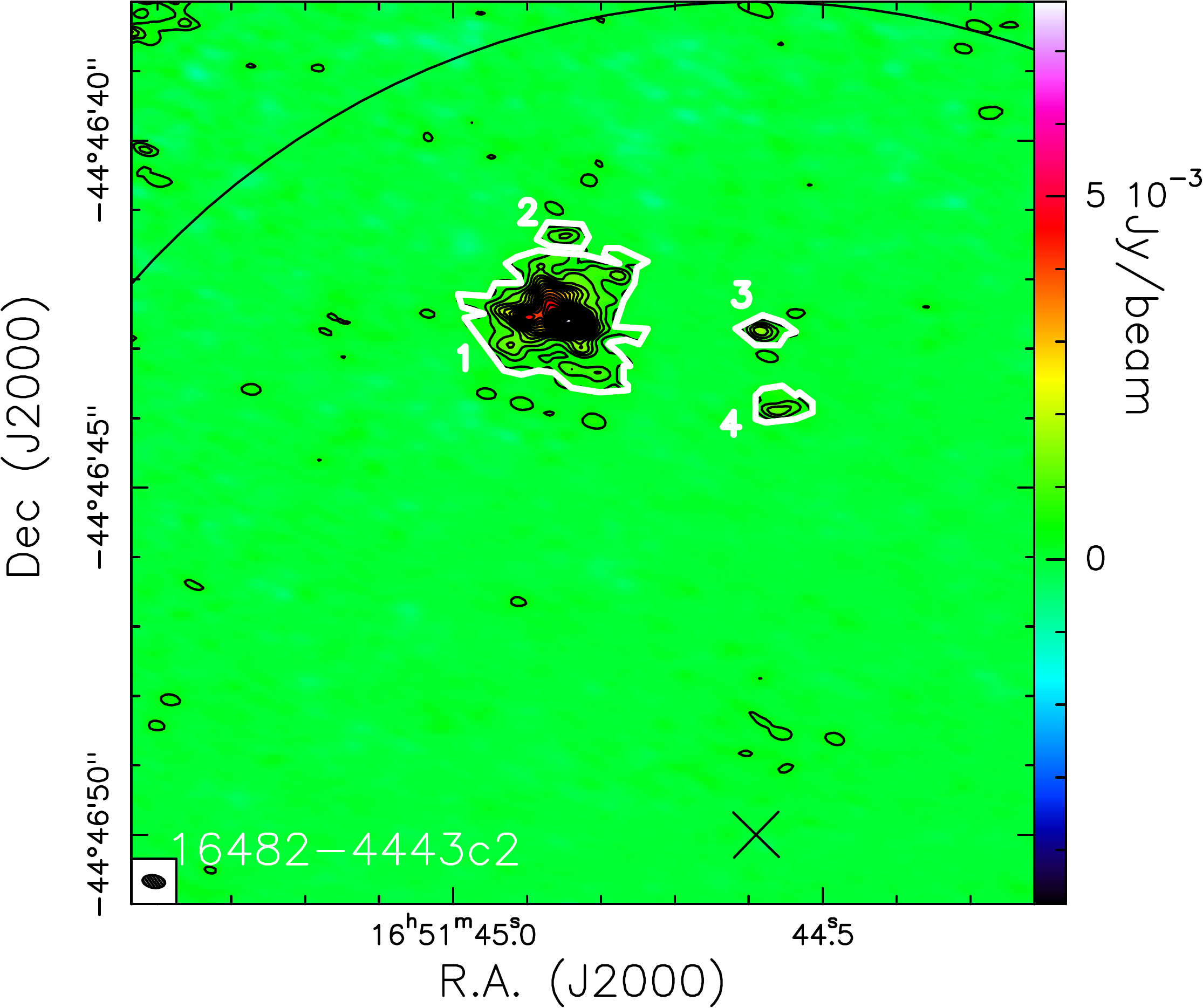}}
\caption{Same as Fig.~\ref{fig_map_08477} for 16482--4443c2. 
The first contour level, and the step, is $3\times 10^{-4}$ Jy beam$^{-1}$, corresponding
to the 3$\sigma$ rms noise level ($1\sigma \sim 1\times 10^{-4}$ Jy beam$^{-1}$).
}
\label{fig_map_16482}
\end{figure}

\begin{figure}[!]
\centerline{\includegraphics[width=9cm,angle=0]{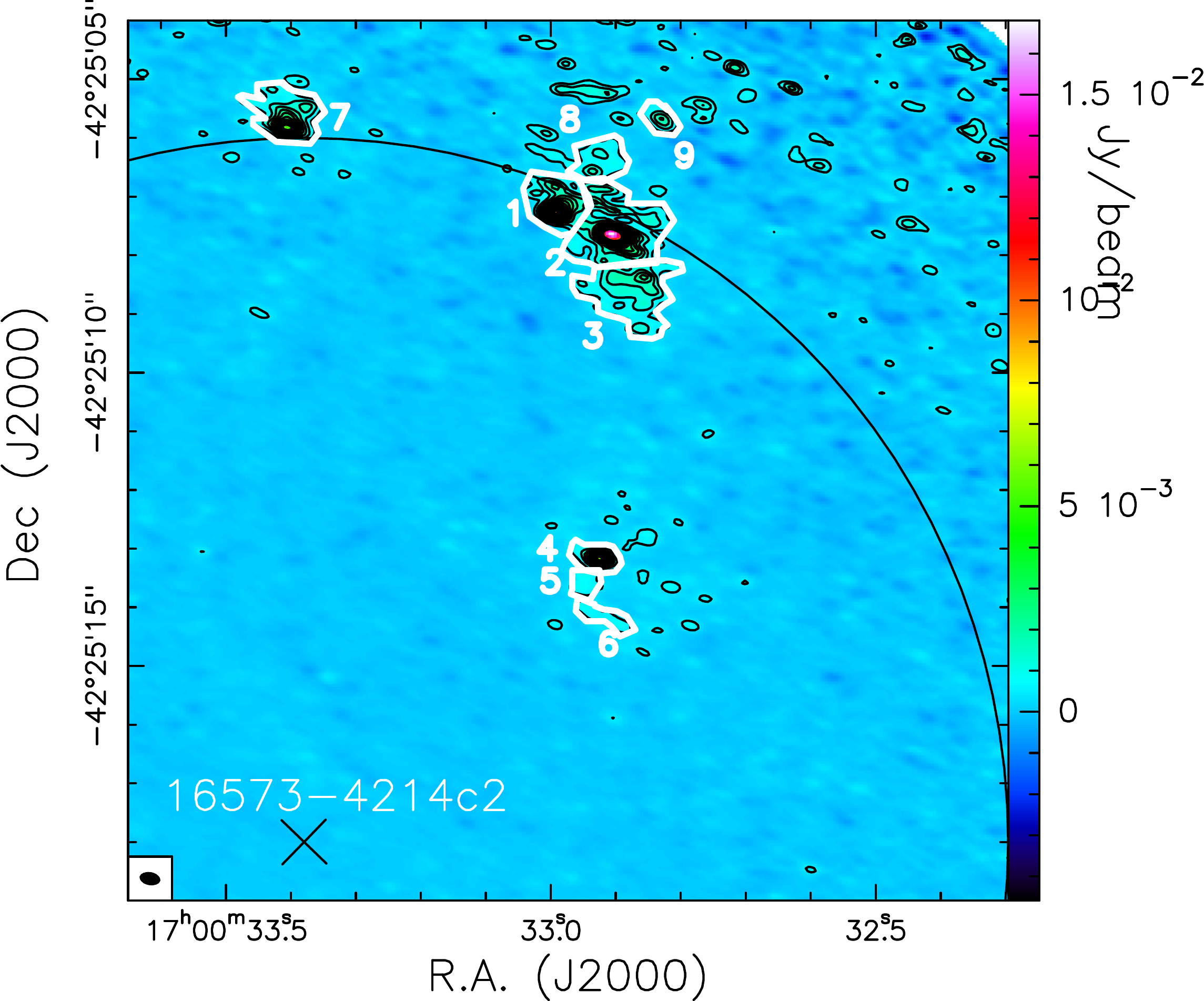}}
\caption{Same as Fig.~\ref{fig_map_08477} for 16573--4214c2. 
The first contour level, and the step, is $4.8\times 10^{-4}$ Jy beam$^{-1}$, corresponding
to the 3$\sigma$ rms noise level ($1\sigma \sim 1.6\times 10^{-4}$ Jy beam$^{-1}$).
}
\label{fig_map_16573}
\end{figure}

In Figs.~\ref{fig_map_08477} to ~\ref{fig_map_16573}, we show the identified 
fragments in each source, while in tables~\ref{tab_08477} to~\ref{tab_16573} we 
list their main properties: peak position, integrated flux density ($F_{\nu}$), peak flux 
density ($F_{\nu}^{\rm peak}$), diameter ($D$), and mass ($m$). 
The coordinates of each fragment indicate the position of its peak flux. The other parameters
have beed derived as explained in Sect.~\ref{res_prop}.
The map of 16061--5048c1 is not shown because already published in Fontani et 
al.~(\citeyear{fontani2016}), following the same approach for the fragment identification.


\begin{table*}
\begin{center}
\caption[] {Peak position (in R.A. and Dec. J2000), integrated flux $F_{\nu}$ (inside the 3$\sigma$ rms 
contour level), peak flux $F_{\nu}^{\rm peak}$, diameter $D$, and mass $m$ of the fragments 
identified in Fig.~\ref{fig_map_08477} towards 08477--4359c1. The average error on $F_{\nu}$
is about $1-2 \times 10^{-3}$ Jy beam$^{-1}$. }
\label{tab_08477}
\begin{tabular}{lcccccc}
\hline \hline
Fragment	& R.A. (J2000) & Dec. (J2000) & $F_{\nu}$ & $F_{\nu}^{\rm peak}$ & $D$ & $m$  \\
      &  h:m:s          &    $^{\circ}$:$^{\prime}$:\asec & mJy          & mJy beam$^{-1}$          & pc    & \solm\ \\
      \hline
1     & 08:49:35.86 &  $-$44:11:55.1            & 39     &     6.11         & 0.018  &    1.47 \\
2     & 08:49:35.85 &  $-$44:11:56.3            & 34     &    19.7          & 0.011  &    1.28 \\
3     & 08:49:35.73 &  $-$44:11:57.1            & 6.6   &    1.40          & 0.009  &    0.24 \\
4     & 08:49:34.74 &  $-$44:11:55.1            & 20     &     2.93         & 0.014  &    0.75 \\
\hline
\hline
\end{tabular}
\end{center}
\end{table*}

\begin{table*}
\begin{center}
\caption[] {Same as Table~\ref{tab_08477} for 15470--5419c1.}
\label{tab_15470c3}
\begin{tabular}{lcccccc}
\hline \hline
Fragment	& R.A. (J2000) & Dec. (J2000) & $F_{\nu}$ & $F_{\nu}^{\rm peak}$ & $D$ & $m$  \\
      &  h:m:s          &    $^{\circ}$:$^{\prime}$:\asec & mJy     & mJy beam$^{-1}$  & pc    & \solm\ \\
      \hline
1     & 15:51:28.83 &   $-$54:31:33.2           &  3.0   & 1.55   & 0.020  &   0.63 \\
2     & 15:51:27.97 &   $-$54:31:43.2           &  1.0  &  0.87   & 0.013  &   0.22 \\
3     & 15:51:27.96 &   $-$54:31:42.2           &  2.0    & 0.94   & 0.018  &   0.43 \\
4     & 15:51:27.98 &   $-$54:31:40.0           &  3.8    & 2.17   & 0.021  &   0.80 \\
5     & 15:51:28.01 &   $-$54:31:38.4           &  2.0    & 1.17   & 0.017  &   0.43 \\
6     & 15:51:27.96 &   $-$54:31:30.5           &  3.3   &  0.85   & 0.022  &   0.70 \\
7     & 15:51:27.92 &   $-$54:31:32.0           &  11      & 4.00   & 0.029  &   2.37 \\
8     & 15:51:27.46 &   $-$54:31:34.8           &  1.5    & 1.02   & 0.016  &   0.33 \\
9     & 15:51:27.38 &   $-$54:31:40.5           &  3.1   & 1.83   & 0.020  &   0.65 \\
10    & 15:51:27.33 &   $-$54:31:35.6           &  0.4    & 0.69   & 0.010  &   0.09 \\
11    & 15:51:27.22 &   $-$54:31:46.6           &  1.0    & 0.63   & 0.014  &   0.20 \\
12    & 15:51:28.61 &   $-$54:31:28.8           &  2.4    & 1.9     & 0.018  &   0.51 \\
13    & 15:51:29.19 &   $-$54:31:31.1           &  57.5  & 10.0   & 0.042  & 12.1 \\
14    & 15:51:29.24 &   $-$54:31:32.1           &  21.8  & 13.0   & 0.026  & 4.60 \\
\hline
\hline
\end{tabular}
\end{center}
\end{table*}

\begin{table*}
\begin{center}
\caption[] {Same as Table~\ref{tab_08477} for 15470--5419c3.}
\label{tab_15557c3}
\begin{tabular}{lcccccc}
\hline \hline
Fragment	& R.A. (J2000) & Dec. (J2000) & $F_{\nu}$ & $F_{\nu}^{\rm peak}$ & $D$ & $m$  \\
      &  h:m:s          &    $^{\circ}$:$^{\prime}$:\asec & mJy     & mJy beam$^{-1}$  & pc    & \solm\ \\
      \hline
1      & 15:51:01.61    &  $-$54:26:39.6 &  4.9        &   1.11  &  0.027   &      0.96 \\
2      & 15:51:01.53    &  $-$54:26:37.8 &  19.9      &   11.5  &  0.033   &      3.89 \\
3      & 15:51:01.51    &  $-$54:26:34.6 &  45.6      &   21.5   &  0.027   &      8.91 \\
4      & 15:51:01.44    &  $-$54:26:35.6 &  37.4      &   9.10   &  0.042   &      7.31 \\
5      & 15:51:01.43    &  $-$54:26:37.3 &  1.7       &   8.4    &  0.017   &     0.34 \\
6      & 15:51:01.29    &  $-$54:26:35.1 &  1.2       &   0.70   &  0.015   &     0.23 \\
7      & 15:51:01.08    &  $-$54:26:46.1 &  0.65     &   0.77   &  0.011   &      0.13 \\
8      & 15:51:00.86    &  $-$54:26:50.4 &  3.6       &   1.38   &  0.021   &      0.71 \\
9      & 15:51:01.43    &  $-$54:26:31.4 &  35.3     &   15.3   &  0.033   &   6.90  \\
\hline
\hline
\end{tabular}
\end{center}
\end{table*}

\begin{table*}
\begin{center}
\caption[] {Same as Table~\ref{tab_08477} for 15557--5215c2.}
\label{tab_15557c2}
\begin{tabular}{lcccccc}
\hline \hline
Fragment	& R.A. (J2000) & Dec. (J2000) & $F_{\nu}$ & $F_{\nu}^{\rm peak}$ & $D$ & $m$  \\
      &  h:m:s          &    $^{\circ}$:$^{\prime}$:\asec & mJy          & mJy beam$^{-1}$  & pc    & \solm\ \\
      \hline
1     & 15:59:36.89     &  $-$52:22:53.9  &  7.6  &    3.1   &  0.023    &     1.32 \\
2     & 15:59:36.88     &  $-$52:22:54.7  &  11     &    5.60  &  0.022   &      1.88 \\
3     & 15:59:36.79     &  $-$52:22:55.5  &  12     &    5.10  &  0.023   &      2.02 \\
4     & 15:59:36.58     &  $-$52:22:55.9  &  22     &    14.0  &  0.025   &      3.87 \\
5     & 15:59:36.53     &  $-$52:22:56.8  &  2.9  &    1.68  &  0.016   &      0.51 \\
6     & 15:59:36.50     &  $-$52:22:52.7  &  54     &    46.0  &  0.025   &      9.45 \\
7     & 15:59:36.41     &  $-$52:22:56.5  &  3.0  &    2.60  &  0.014   &      0.52 \\
8     & 15:59:36.07     &  $-$52:22:55.3  &  12     &    6.1  &  0.024   &      2.07 \\
9     & 15:59:36.14     &  $-$52:22:55.6  &  1.6  &    1.6  &  0.012    &    0.28 \\
10    & 15:59:36.00     &  $-$52:22:58.9  &  2.2  &    1.4  &  0.015   &     0.39 \\
11    & 15:59:36.03     &  $-$52:22:52.1  &  1.7  &    1.3  &  0.013   &     0.30 \\
12    & 15:59:35.93     &  $-$52:22:52.4  &  1.4  &    1.4  &  0.011   &     0.19 \\
\hline
\hline
\end{tabular}
\end{center}
\end{table*}

\begin{table*}
\begin{center}
\caption[] {Same as Table~\ref{tab_08477} for 16061--5048c4.}
\label{tab_16061c4}
\begin{tabular}{lcccccc}
\hline \hline
Fragment	& R.A. (J2000) & Dec. (J2000) & $F_{\nu}$ & $F_{\nu}^{\rm peak}$ & $D$ & $m$  \\
      &  h:m:s          &    $^{\circ}$:$^{\prime}$:\asec & mJy          & mJy beam$^{-1}$  & pc    & \solm\ \\
      \hline
1      & 16:10:06.20    &  $-$50:57:11.9 &   7.24  &   0.60   &  0.023    &      1.92 \\
2      & 16:10:06.07    &  $-$50:57:11.3 &   8.80  &   2.90   &  0.017   &      2.33 \\
3      & 16:10:06.10    &  $-$50:57:12.4 &   0.69  &   0.52   &  0.007  &      0.18 \\
4      & 16:10:05.93    &  $-$50:57:13.1 &   0.93  &   0.68   &  0.008  &      0.25 \\
\hline
\hline
\end{tabular}
\end{center}
\end{table*}

\begin{table*}
\begin{center}
\caption[] {Same as Table~\ref{tab_08477} for 16482--4443c2.}
\label{tab_16482}
\begin{tabular}{lcccccc}
\hline \hline
Fragment	& R.A. (J2000) & Dec. (J2000) & $F_{\nu}$ & $F_{\nu}^{\rm peak}$ & $D$ & $m$  \\
      &  h:m:s          &    $^{\circ}$:$^{\prime}$:\asec & mJy          & mJy beam$^{-1}$  & pc    & \solm\ \\
      \hline
1      &  16:51:44.85  &  $-$44:46:42.6 &     69.4   &  7.68  &  0.038    &     14.12  \\
2      &  16:51:44.85  &  $-$44:46:41.4 &     1.64   &  1.05 &  0.009   &     0.33 \\
3      &  16:51:44.58  &  $-$44:46:42.8 &     2.27   &  1.63  &  0.010   &     0.46 \\
4      &  16:51:44.56  &  $-$44:46:43.9 &     2.81  &  1.22  &  0.012    &     0.57 \\
\hline
\hline
\end{tabular}
\end{center}
\end{table*}

\begin{table*}
\begin{center}
\caption[] {Same as Table~\ref{tab_08477} for 16573--4214c2.}
\label{tab_16573}
\begin{tabular}{lcccccc}
\hline \hline
Fragment	& R.A. (J2000) & Dec. (J2000) & $F_{\nu}$ & $F_{\nu}^{\rm peak}$ & $D$ & $m$  \\
       &  h:m:s           &    $^{\circ}$:$^{\prime}$:\asec & mJy          & mJy beam$^{-1}$  & pc    & \solm\ \\
       \hline
1      & 17:00:32.99  &    $-$42:25:07.3 &    21.3      &   9.72 &  0.012      &    1.96 \\
2      & 17:00:32.91  &    $-$42:25:07.6 &    39.0      &   17.5 &  0.015     &    3.59 \\
3      & 17:00:32.86  &    $-$42:25:08.4 &    21.7      &   2.42 &  0.017     &    2.00 \\
4      & 17:00:32.92  &    $-$42:25:13.2 &    7.88      &   5.70 &  0.009    &    0.73 \\
5      & 17:00:32.95  &    $-$42:25:13.6 &    1.93      &   0.85 &  0.006    &    0.18 \\
6      & 17:00:32.94  &    $-$42:25:14.1 &    2.33      &   0.84 &  0.007    &    0.21 \\
7      & 17:00:33.41  &    $-$42:25:05.9 &    18.6      &   5.10 &  0.014    &   1.71  \\
8      & 17:00:32.94  &    $-$42:25:06.6 &    6.20      &   1.60 &  0.011    &   0.57  \\
9      & 17:00:32.83  &    $-$42:25:05.7 &    3.25      &   1.80 &  0.007   &   0.30  \\
\hline
\hline
\end{tabular}
\end{center}
\end{table*}

\clearpage

\renewcommand{\thetable}{B-\arabic{table}}
\renewcommand{\thefigure}{B-\arabic{figure}}
\renewcommand{\thesection}{B-\arabic{section}}
\setcounter{table}{0}
\setcounter{figure}{0}
\setcounter{section}{0}
\section*{Appendix B: Sink particles distribution in simulations}
\label{appB}

\begin{figure*}[htp]
\includegraphics[width=18cm,angle=0]{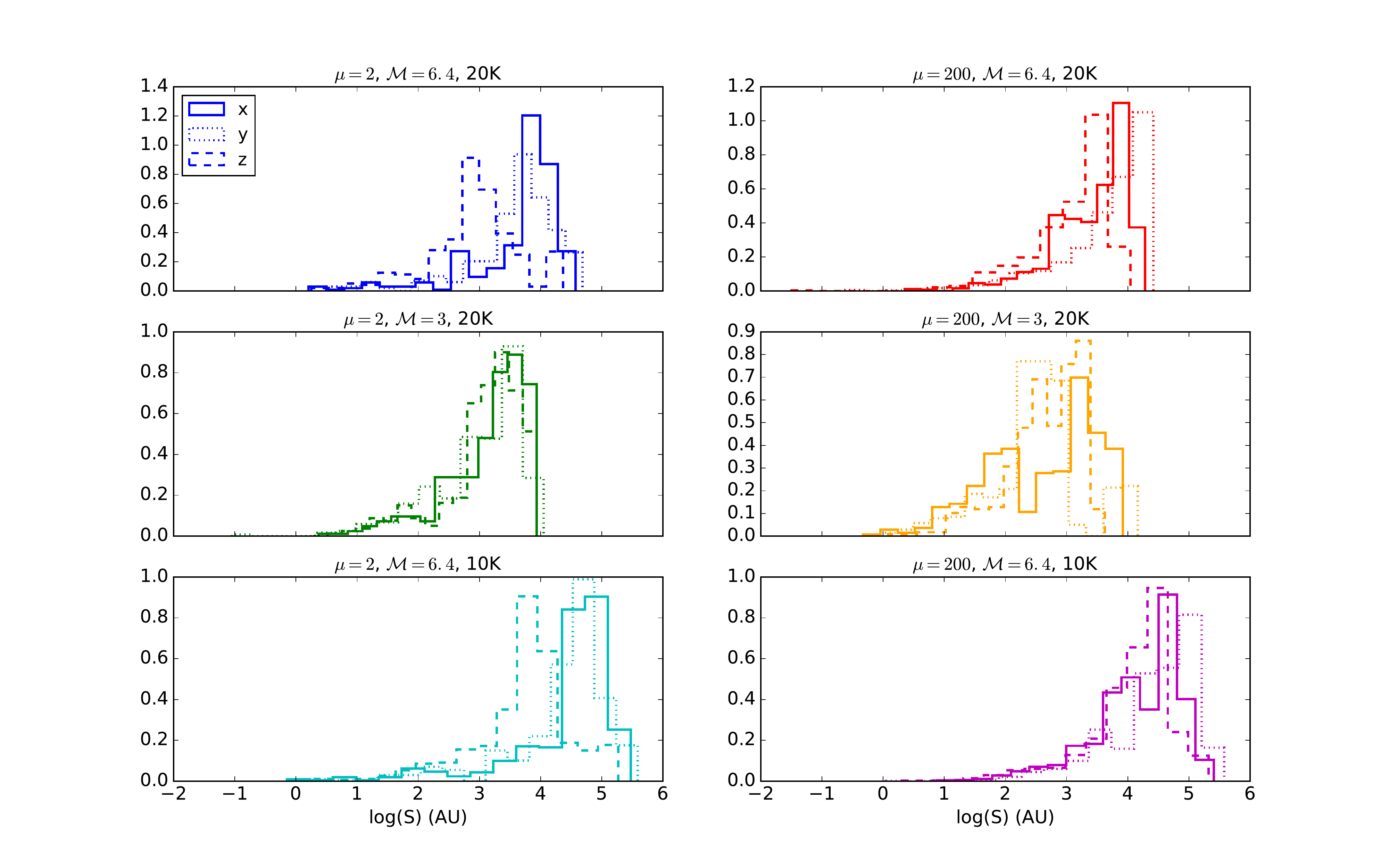}
\caption{Histograms of the sink particles separation distribution at a SFE of 15\%. The top row shows simulations (2), middle simulations (3), and bottom simulations (4). The left (resp. right) column shows the $\mu=2$ ($\mu=200$) cases. The solid line represents the separation distribution in the $x$-direction, the dotted line that in the $y$-direction, and the dashed line that in the $z$-direction.}
\label{fig:sep}
\end{figure*}

Figure  \ref{fig:sep} shows the histograms of the separation distribution for simulations (2), (3), and (4) along the three coordinates axis. The trend of figure~\ref{time:evol} is recovered: the largest separation are found for simulations (4) (middle row) while the smallest one are found in simulations (3) (bottom row). In the $(\mu=2,\mathcal{M} \sim 6.4)$, the separation in the $z$-direction is smaller than in the other two directions, with a difference of a factor $\sim 10$. This means that a filamentary structure with an aspect ratio of 1/10 can be seen by looking at the sink particle distribution in two directions. This feature is only present the strongly magnetized and most turbulent simulations. In all other case, we find a more compact size distribution, suggesting a more roundish sink particle distribution.

\begin{figure*}[htp]
\includegraphics[width=18cm,angle=0]{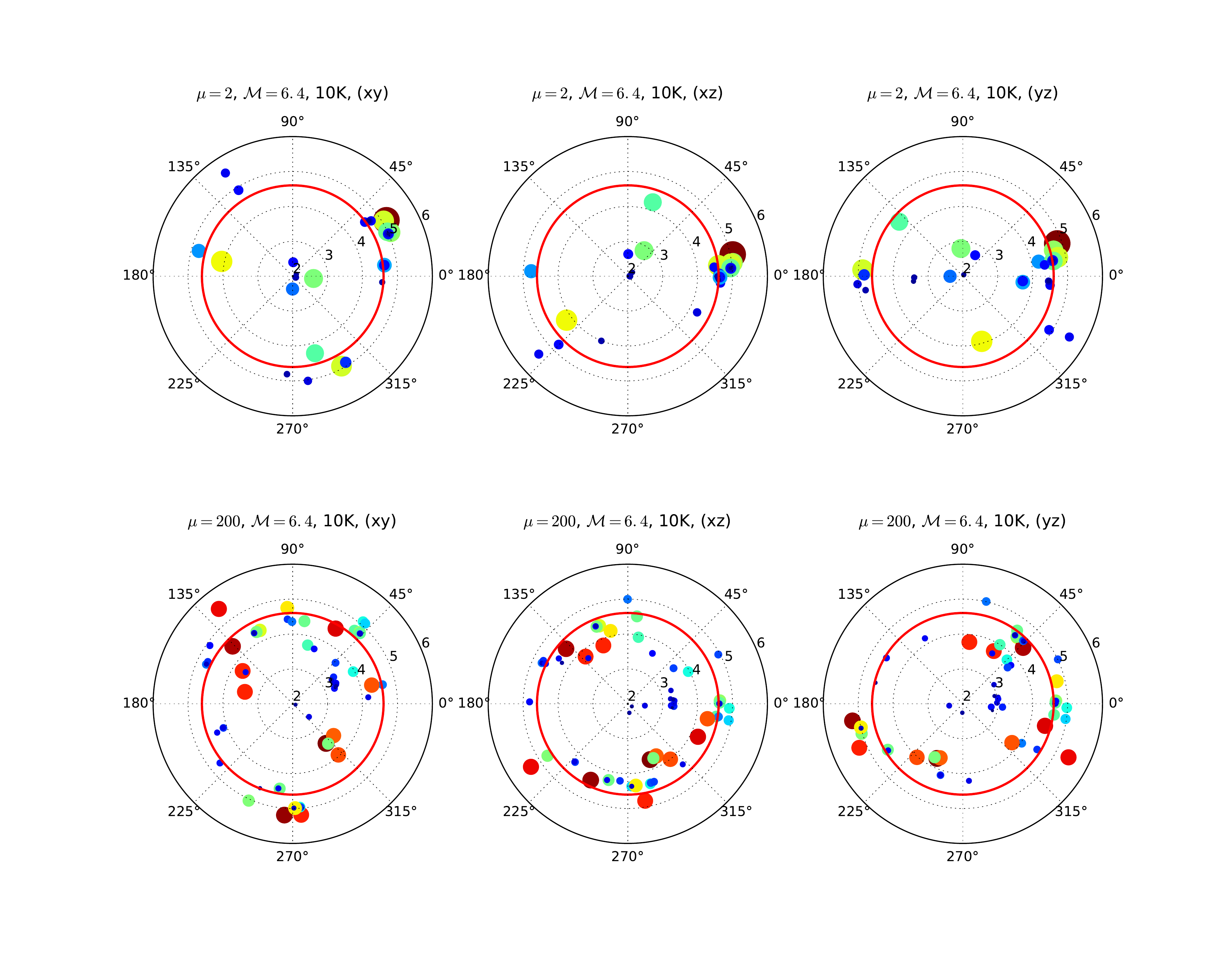}
\caption{Projected sink particle distribution centered around the most massive sink particles at an SFE of 15\% for simulations (3)  ($\mathcal{M}\sim 6.4$, $T=10$~K). The radial direction shows the distance in au in logarithmic scale. The red circle represents the size of the region that we post-processed with CASA to produce the ALMA synthetic observations. }
\label{fig:sep_proj}
\end{figure*}

Figure \ref{fig:sep_proj} shows the 2D projected sink particles distribution around the most massive one for simulations (3). The red circle delimits the region within a radius of 40000~au that would be observed in our ALMA synthetic observations. A non-negligible number of sink particles is thus excluded from analaysis, and would not be picked up by ALMA in the configuration we used. 

\end{document}